\documentclass[aip,pop,article-title,reprint]{revtex4-1}
\usepackage{hyperref}
\usepackage{graphicx}
\usepackage{amsmath}
\usepackage{amssymb}
\usepackage{bm}
\def\R{\bm R}
\def\x{\bm x}
\def\z{\bm z}
\def\X{\bm X}
\def\B{\bm B}
\def\v{\bm v}
\def\E{\bm E}
\def\r{\bm r}

\begin{document}
\title{Finite gyro-radius multidimensional electron hole equilibria}
\author{I H Hutchinson}
\email{ihutch@mit.edu}
\affiliation{Plasma Science and Fusion Center\\
Massachusetts Institute of Technology\\
Cambridge, MA 02139, USA}

\begin{abstract}
  Finite electron gyro-radius influences on the trapping and charge
  density distribution of electron holes of limited transverse extent
  are calculated analytically and explored by numerical orbit
  integration in low to moderate magnetic fields. Parallel trapping is
  shown to depend upon the gyro-averaged potential energy and to give
  rise to gyro-averaged charge deficit. Both types of average are
  expressible as convolutions with perpendicular Gaussians of width
  equal to the thermal gyro-radius. Orbit-following confirms these
  phenomena but also confirms for the first time in self-consistent
  potential profiles the importance of gyro-bounce-resonance
  detrapping and consequent velocity diffusion on stochastic orbits.
  The averaging strongly reduces the trapped electron deficit that can
  be sustained by any potential profile whose transverse width is
  comparable to the gyro-radius $r_g$. It effectively prevents
  equilibrium widths smaller than $\sim r_g$ for times longer than a
  quarter parallel-bounce-period. Avoiding gyro-bounce resonance
  detrapping is even more restrictive, except for very small potential
  amplitudes, but it takes multiple bounce-periods to
  act. Quantitative criteria are given for both types of orbit loss.
  \end{abstract}
\maketitle

\section{Introduction}

Solitary potential peaks, of extent a few Debye-lengths parallel to
the ambient magnetic field, are frequently observed by satellites in
space
plasmas\cite{Matsumoto1994,Ergun1998,Bale1998,Mangeney1999,Pickett2008,Andersson2009,Wilson2010,Malaspina2013,Malaspina2014,Vasko2015,Mozer2016,Hutchinson2018b,Mozer2018}. They
are usually interpreted as being electron holes: a type of nonlinear
Bernstein, Greene, Kruskal [BGK] Vlasov-Poisson
equilibrium\cite{Bernstein1957,Hutchinson2017} in which a deficit of
electrons on trapped orbits causes the positive charge density that
sustains them.  Such structures are frequently observed to form in
non-linear one-dimensional particle simulations of unstable electron
distribution functions of types such as two-stream or
bump-on-tail\cite{Mottez1997,Miyake1998a,Goldman1999,Oppenheim1999,Muschietti2000,Oppenheim2001b,Singh2001,Lu2008}. One-dimensional
dynamical analysis is, however, not sufficient to describe naturally
occurring holes fully, because their transverse dimensions are
limited. The observational evidence, backed up by simulations in a
variety of contexts, is that while holes are generally oblate, in
other words more extended in the perpendicular than in the parallel
(to $B$) direction, their aspect ratio can be as low as
$L_\perp/L_\parallel\sim 1$. Moreover one-dimensional holes have been
shown
analytically\cite{Hutchinson2018a,Hutchinson2019,Hutchinson2019a} and
computationally\cite{Muschietti2000,Wu2010} to be unstable to
perturbations of finite transverse wavelength, which cause them
quickly to break up into multidimensional structures unless the
magnetic field ($B$) is very strong. Multi-satelite missions (e.g.\
Cluster\cite{Graham2016} and Magnetosphere
Multiscale\cite{Steinvall2019,Lotekar2020}) are now beginning to
document the transverse potential structure of electron holes in
space.

The present work culminates a series of theoretical studies addressing
for multidimensional electron hole equilibria the important effects of
finite transverse size. Beyond the one-dimensional BGK treatment that
has dominated past analysis, the phenomena that need to be understood
and accounted for can be listed as (1) transverse electric field divergence, (2)
supposed gyrokinetic modification of Poisson's equation, (3) orbit
detrapping by gyro-bounce resonance, and (4) gyro-averaging of the
potential and density deficit.

The first (transverse divergence) has been theoretically investigated
for a long
time\cite{Schamel1979,Chen2002,Muschietti2002,Chen2004,Krasovsky2004a,Krasovsky2004},
but predominantly supposing that the potential shape is separable of
the form $\phi_z(z)\phi_r(r)$, where $z$ is the parallel and $r$ the
transverse coordinate. In a recent paper in this
series\cite{Hutchinson2021a} (which should be consulted for a review
of prior multidimensional equilibrium studies) it was shown that
electron holes that are solitary can never be exactly separable, and
shown how to construct fully self-consistent potential electron holes
synthetically. Transverse divergence is present even in the limit of
small gyro-radius, and so this phenomenon has been treated in the
context of purely one-dimensional electron motion.

Effect (2) (gyrokinetic polarization modification of Poisson's
equation) was hypothesized as an explanation of statistical
observations that electron hole transverse scale (oblateness)
increases with gyro-radius, or inverse magnetic field
strength\cite{Franz2000}. And it has often been invoked
since\cite{Vasko2017,Vasko2018,Holmes2018,Tong2018,Fu2020}. However,
it is based on a misunderstanding of gyrokinetic theory, as has been
explained and demonstrated in recent work in this
series\cite{Hutchinson2021}. The anisotropic shielding phenomenon
envisaged does not occur and so cannot explain electron hole
aspect-ratio trends.

Phenomena of type (3), gyro-bounce resonance, lead to islands in
trapped phase-space where the gyro-frequency is an even harmonic of
the parallel bounce frequency of electrons. Orbits involving
overlapped islands become stochastic. They start trapped with negative
parallel energy, but energy is resonantly transferred from
perpendicular to parallel velocity causing them to become untrapped
after some moderate number of bounces. Corresponding effects cause
initially untrapped orbits to become trapped. Although this is a
finite gyro-radius effect, it can be treated in a small gyro-radius
approximation that accounts only for parallel motion. An analytic
treatment then proves to be feasible and agrees extremely well with
full numerical orbit integration in a potential whose local radial
potential gradient is specified. The analysis, presented in a
paper in this series\cite{Hutchinson2020}, shows that stochastic orbits, giving rise to
effective velocity-space diffusion, begin (and are always present to
some degree) at the phase-space boundary between trapped and untrapped
orbits (zero parallel energy). Increasing transverse electric field
or, more importantly, decreasing magnetic field causes the energy
range over which orbits are stochastic to extend downward into the
bulk of the trapped orbits.  Depletion of trapped phase-space sets in
rapidly when low harmonic resonances are present. They are, when
$\Omega/(\omega_p\sqrt{\psi})\lesssim 2$ (where $\Omega$ is the
electron gyro-frequency, $\omega_p$ the plasma frequency, and $\psi$
the peak electron hole potential in units of electron temperature
$T_e/e$). Thus $\Omega/(\omega_p\sqrt{\psi})\gtrsim 2$ is an effective
threshold for the existence of long-lived multidimensional electron
hole equilibria. That prior work used a purely nominal locally
exponential radial potential profile. The present work uses full
equilibrium potential structures of plausible shape. 

Phenomenon (4), gyro-averaging, is the main topic of the present work,
taking fully into account gyro-radii that are \emph{not} small
relative to the perpendicular scale length. It is discussed
analytically in section \ref{sec2}, and validated by numerical orbit
integration in section \ref{sec3}, which includes exploration of
resonant detrapping over the entire radial profile of axisymmetric
holes whose velocity distribution functions are constrained by the
effective diffusivity of detrapping regions. In section \ref{sec4} the
consequences of detrapping for the feasible shape and parameters of
electron holes are discussed

\begin{figure}
  \centering
  \includegraphics[width=0.9\hsize]{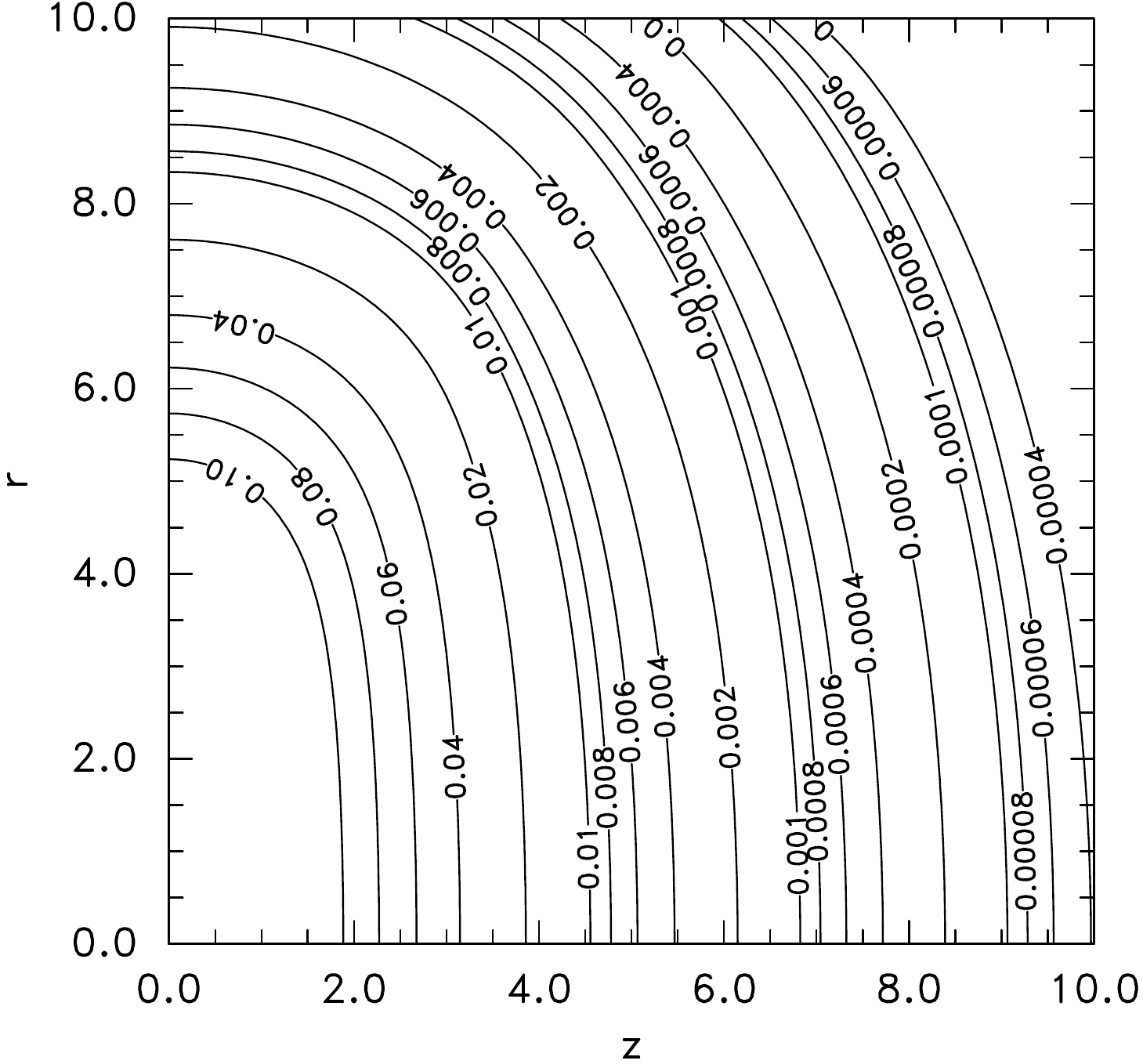}
  \caption{Example electron hole shape, showing contours of potential
    $\phi(r,z)$. Only a quadrant of the structure is shown, it being
    reflectionally symmetric about the $r$ and $z$ axes.}
  \label{fig:phiofrz1}
\end{figure}

The treatment is, for specificity, of axisymmetric potential
structures, independent of the angle $\theta$ of a cylindrical
$r,z,\theta$ coordinate system. Fig.\ \ref{fig:phiofrz1} illustrates
an example.
However, most of the gyro-averaging
effects are immediately generalizable to non-axisymmetric potentials,
and the drift orbits are always normal to the potential gradient.
Throughout this discussion, ions are considered an immobile
neutralizing background charge density and excluded from the
analysis. This is justified when the electron hole moves (in the
parallel direction) at speeds much exceeding the ion thermal or sound
speeds, because ions' much greater mass gives them much slower
response than electrons. Electron holes can travel at any speed from a
few times the ion sound speed up to the electron thermal
speed. Therefore for the vast majority of this speed range ignoring
ion response is a good approximation.

\section{Gyro-averaging analytics}
\label{sec2}

Phenomena (1) and (3) can be treated using a pure drift kinetic
treatment, taking the electrons to be sufficiently localized in
transverse position not to require finite gyro-radius to be accounted
for in trapping and bounce motion. When, however, the gyro-radius is
not negligible compared with the scale-length of transverse variation,
the effects of varying potential and parallel electric field around
the gyro-orbit must be considered.  This gyro-averaging (rather than
in the mistaken supposition (2)) is where the sorts of techniques
familiar in gyrokinetics become influential.

There are actually two conceptually different ways to express the
``center'' of an approximately helical gyro orbit of a particle in
magnetic and electric fields. One is to define it in the stationary
reference frame as $\R=\x+\v \times \hat\z/\Omega$, where $\x$ and
$\v$ are the instantaneous position and velocity of a particle, and
$\Omega=qB/m$ is the (cyclotron- or) gyro-frequency in the magnetic
field (here taken uniform) $\B=B\hat\z$. This definition is commonly
used in gyrokinetics derivations. The other, which was associated with
the original development of orbit drift treatments
(see e.g.\ Northrop (1961)\cite{Northrop1961}), and is therefore historically the \emph{first}
definition, is to regard it as the position of the particle averaged
over the gyration, in the frame moving with the mean drift
velocity. This \emph{first} definition was originally, and will be
here called the ``guiding-center''. If the (reference frame)
perpendicular drift velocity is $\v_d$ then the particle velocity in
the drifting frame is $\v-\v_d$ and the guiding-center is at
$\X=\x+(\v-\v_d)\times \hat\z/\Omega=\R-\v_d\times \hat\z/\Omega$
about which the motion of the particle is approximately a centered
circle. These two definitions therefore differ by a distance
$v_d/\Omega$ normal to the direction of $v_d$. With uniform $\B$, the
drift is $\v_d=\E\times \B /B^2$; so $\X-\R=\E_\perp /B\Omega$. This
difference is the integrated polarization drift, or simply the
``polarization'', that would arise by increasing the electric field
from zero to $\E_\perp$. Because the reference frame definition
``gyro-center'' ($\R$) is actually \emph{not} at the center of the
circle which the orbit follows in the drift frame, when an orbit
average is carried out relative to $\R$, then an extra term
representing the polarization $\X-\R$ must be included. It accounts
for the gyro-radius distance from $\R$ to the particle \emph{varying
  with phase angle} around the orbit. It is more convenient for our
current purposes to adopt the \emph{guiding-center} ($\X$) as our
reference position, because the gyro-radius (guiding-center to
particle) is then (to relevant order) independent of gyro-angle; and
polarization drift is irrelevant. Naturally a thorough mathematical
treatment (which is not the present purpose) will get the same results
from either perspective. We will henceforward denote the
\emph{guiding-center} as $\x_c$ and the \emph{gyro-radius} relative to
it as $\r_g$.

Now consider a finite gyro-radius particle orbit that moves along the
magnetic field. We are most interested in particles whose total
energy, $W=mv^2/2+q\phi$, is positive because when it is negative,
this exactly conserved energy guarantees trapping. To give sufficient
sustaining charge deficit, electron holes require extensive trapping
of orbits with positive $W$.  Whether or not such an electron is
trapped depends on whether the parallel electric field attracting it
to $z=0$ is sufficient, when integrated over the orbit, to reduce its
parallel momentum to zero and hence cause it to bounce. In the absence
of gyration, this condition would be deduced from
$m \dot v_\parallel =qE_\parallel$, in the integrated form
$[mv_\parallel^2/2]=\int m {d\over dt}(v_\parallel^2)dt=-q\int
{d\phi\over dz} dz= -[q\phi]$ (parallel energy conservation) and requires
$W_\parallel \equiv mv_\parallel^2/2+q\phi<0$. Clearly, in the
presence of gyration, we must regard the potential (difference) as
involving instead the potential gradient \emph{averaged over the orbit},
including the variation arising from gyro-motion when $\phi$ varies in
the perpendicular coordinate. This is a (finite-duration)
gyro-average in which end-effects introduce some possible variation
with initial gyro-angle (e.g.\ at $z=0$: the plane of reflectional
symmetry of the hole and maximum potential along lines of constant
$r$). But provided sufficiently many gyro-periods elapse in moving to
distant $z$ where $\phi\to 0$, and provided the potential variation is
smooth enough that there is no other substantial cause of transfer of
kinetic energy from perpendicular to parallel motion, it will be a good
approximation to consider the effective potential drop between $z=0$
and $z=\infty$ to be given by the gyro-averaged potential at $z=0$:
\begin{equation}\label{barphi}
\bar \phi(\x_{c\perp},0)\equiv\int_0^{2\pi}
\phi(x_c+r_g\cos\xi,y_c+r_g\sin\xi,0)\;{d\xi\over 2\pi}.
\end{equation}
This potential gyro-average about the guiding-center is approximately
what determines particle trapping. An orbit is (provisionally)
trapped if $-q\bar\phi >mv_\parallel^2/2$ at $z=0$.

A second gyro-average is required to derive the particle density
(which is what determines the charge density in Poisson's equation)
from the guiding-center density. At any position $\x$, contributions
to the particle density arise from all guiding-center positions $\x_c$
on a circle a perpendicular distance $r_g$ away. Consequently if the
distribution function of guiding-centers is
$f_c(\x_c,v_\perp,v_\parallel)$, and we write the vector
$\r_g=(\cos\xi,\sin\xi,0)r_g$, the corresponding distribution of
particles is
\begin{equation}
  \label{gcp}
  \begin{split}
  f(\x,v_\perp,v_\parallel)
  &=\int_0^{2\pi}f_c(\x_c,v_\perp,v_\parallel)\; {d\xi\over 2\pi}\\
  &=\int_0^{2\pi}f_c(\x-\r_g,v_\perp,v_\parallel)\; {d\xi\over 2\pi}.
\end{split}
\end{equation}
[If there is an electric field gradient $\nabla_\perp \E_\perp$, and
hence drift gradient $\nabla_\perp \v_d$, then if $r_g$ is taken
constant, $v_\perp$ varies with gyroangle $\xi$ in
$f$. This higher order effect can be ignored for present purposes.]

\subsection{Gyro-averaged potential}

In standard gyrokinetic theory, the evaluation of the gyro-averages
usually relies upon a Fourier transform of the perpendicular potential
variation. When $\phi(\x)=\phi(k_x){\rm e}^{ik_x x}$, its gyro-average
is proportional to a Bessel function
\begin{equation}
  \label{fourier}
  \bar\phi =\phi(k_x)\int {\rm e}^{ik_x r_g \cos\xi}\;
  {d\xi\over 2\pi}=J_0(k_xr_g) \phi(k_x). 
\end{equation}
In gyrokinetic theory, there is a perturbation in the guiding-center
distribution function ($f_c$) that is proportional to the
gyro-averaged potential. This perturbation arises from the integration
along orbits of the Vlasov equation (from a distant/past equilibrium).
To find the particle distribution function, needed for current- and
charge-density (Poisson's equation in the electrostatic limit) then
requires a second gyro-average, eq.\ (\ref{gcp}), which introduces
another $J_0(k_xr_g)$ factor.  The double gyro-average and integration
over a Maxwellian $v_\perp$ distribution then gives rise to a modified
Bessel function form proportional to
${\rm e}^{-r_{gt}^2} I_0(r_{gt}^2)$, where $r_{gt}$ is the thermal
gyroradius $v_t/\Omega$.

For trapped orbits in an electron hole equilibrium, the distribution
function cannot feasibly be derived from a distant/past thermal equilibrium. A
trapped particle deficit (relative to reference distribution) arises
by highly complicated non-linear processes leading up to the hole
formation, and is then ``frozen'' on trapped orbits. It is therefore
not appropriate to suppose that one can calculate the trapped particle
distribution function. It must be specified in a form constrained by
plausibility and non-negativity, and be expressed as a function of the
constants of motion: total energy and (approximately) magnetic moment;
so that it is consistent with an eventually steady Vlasov hole
equilibrium.  In any case, for an electron hole the detailed
transverse velocity dependence of the distribution is not a critical
part of our interest.  It is mathematically convenient to adopt a
separable Maxwellian $v_\perp$ form, so that
$f_c=f_{c\parallel}(v_\parallel).{\rm e}^{-(v_\perp/v_t)^2/2}/2\pi
v_t^2$ (where $v_t=\sqrt{T/m}$). The plausibility of this separable
form requires us to suppose that there is not a strong cross-coupling
between parallel and perpendicular velocity that compromises
separation. Since the gyro-averaging process means that parallel
trapping \emph{does} depend to some extent on $v_\perp$ as well as
$v_\parallel$, the separable form will be poorly justified unless the
phase-space density deficit $\tilde f_c$ is small near the
trapped-passing boundary for the majority of the $v_\perp$
distribution. Fortunately $\tilde f_c$ being negligible near
$W_\parallel=0$ is one of the plausibility constraints we must enforce
anyway, to account for orbit stochasticity there.

With this caveat we can perform the integration over $v_\perp$ and
gyro-angle $\xi$ simultaneously as follows, for a single
guiding-center position.
\begin{equation}
  \label{xivave}
  \begin{split}
  \bar\phi(\x_c) &= \int \phi(\x_c+\r_g){\rm e}^{-(v_\perp/v_t)^2/2}
  {v_\perp dv_\perp d\xi\over 2\pi v_t^2}\\
  &=\int\phi(\x_c+\r_g){\rm e}^{-(\r_{g}/r_{gt})^2/2}\;
  {d^2\r_g\over 2\pi r_{gt}^2}.
  \end{split}
\end{equation}
This shows that the simultaneous gyro-averaging and perpendicular
velocity integration simply convolves the quantity of interest
(potential in this case) with a two-dimensional (unit area) Gaussian
transverse profile, of width $r_{gt}\equiv v_t/\Omega$.
Thus, the effective trapped particle confining potential is modified
by finite gyroradius Gaussian smoothing in the transverse
direction. This smoothing will have the effect of decreasing the
height of the effective potential peaks (regions of negative $d^2\phi/d
r^2$) and increasing the height of potential troughs (regions of positive 
$d^2\phi/dr^2$). Trapping is reduced near the origin and enhanced
in the wings, for a domed potential distribution.

\subsection{Gyro-averaged density deficit}

The hole potential is sustained by a deficit $\tilde n$ in trapped
electron density near $z=0$, which gives rise to positive charge
density there.  \iffalse Poisson's equation
$\nabla^2\phi=-\rho/\epsilon_0=(n_\infty-n_f-\tilde n)$ and the form
of $\tilde n(r,\phi)$ determine the potential $\phi(r,z)$. Since the
reference density $n_f$ is approximately $n_\infty(1+\phi)$ the
screening charge density is linear in $\phi(r)$ and hence so is
$\tilde n(r)$. It is therefore plausible to suppose that \else With a
similar caveat about separability of the transverse velocity
distribution function, \fi the gyro-averaging of the trapped
phase-space-density deficit $\tilde f$ in the back-transformation from
$\tilde f_c$ gives rise to a density deficit smoothed in
transverse position in the same way as the potential. That is,
\begin{equation}
  \label{ntsmooth}
  \tilde n   =\int\tilde n_c(\x-\r_g){\rm e}^{-(\r_{g}/r_{gt})^2/2}\;
  {d^2\r_g\over 2\pi r_{gt}^2},
\end{equation}
where $\tilde n_c(\x_c)$ is the deficit in the guiding-center density
($\tilde f_c$ integrated over velocity).

Suppose we compare a situation of finite $r_g$ with a high magnetic
field case: such that $r_{g1}\simeq 0$ purely one-dimensional motion
(subscript 1). If $\tilde n_1(r,\phi)$ and $\tilde n_c(r,\bar\phi)$
were taken to be the same function
[$\tilde n_1(r,\phi)=\tilde n_c(r,\phi)$], except with different
argument ($\phi$ versus $\bar\phi$), then for any given potential
shape $\psi(r)$, the particle density deficit $\tilde n$ at finite
gyro-radius would differ from $\tilde n_1$ by \emph{two} Gaussian
convolutions of this form. One represents the effect of gyro-averaging
of the potential, and the other the transformation from guiding-center
density to particle density. Such $\tilde n$ would not then be
consistent with Poisson's equation. Instead, to keep the $\psi(r)$
fixed, $\tilde n_c(r,\phi)$ would have to be more peaked in $r$ by two
``deconvolutions'' than $\tilde n_1(r,\phi)$ so as to give
$\tilde n=\tilde n_1$. Alternatively, if one insists that $\tilde n_1$
and $\tilde n_c$ must be the same function, then the radial potential
variation $\psi(r)$ of the two cases cannot be the same but must have
$\psi(r)$ smoothed by two Gaussian convolutions relative to
$\psi_1(r)$. Two Gaussian convolutions of the same width $r_{gt}$ are
equivalent to one convolution of width $\sqrt{2}r_{gt}$.

\section{Validation by direct orbit integration }
\label{orbitint}\label{sec3}
Since our discussion has alerted us to the fact that a gyro-averaged
treatment is only an approximation in respect of trapping, it is worth
exploring how good an approximation it is and what phenomena
compromise its accuracy. This has been done by performing full (6-D)
orbit numerical integrations using an updated version of the code
described in reference\cite{Hutchinson2020}. The main upgrades consist
of ability to use any axisymmetric hole potential profile provided
from a different code in the form of an input file, and the
calculation and plotting of various gyro-averaged quantities. In the
code and this and the following section we work in \emph{normalized
  units}: length in Debye-lengths $\lambda_D$, time in inverse plasma
frequencies $1/\omega_p$ and energies in $T_e/e$, for which the
electron charge is then -1 and velocity units are
$v_t=\sqrt{T_e/m}$. The convention throughout is that the value of
potential $\phi$ at position $z=0$ is written $\psi\ [=\phi(r,0)]$,
and the value of $\psi$ at $r=0$ is $\psi_0$ which is the potential at
the origin. Thus $\bar\psi$ is the gyro-average of potential at $z=0$,
but $\bar\phi$ is the average over the orbit at the instantaeous
position: $\phi(r,z)$.

An example of an orbit is shown
\begin{figure}
  \centering
  \includegraphics[width=0.8\hsize]{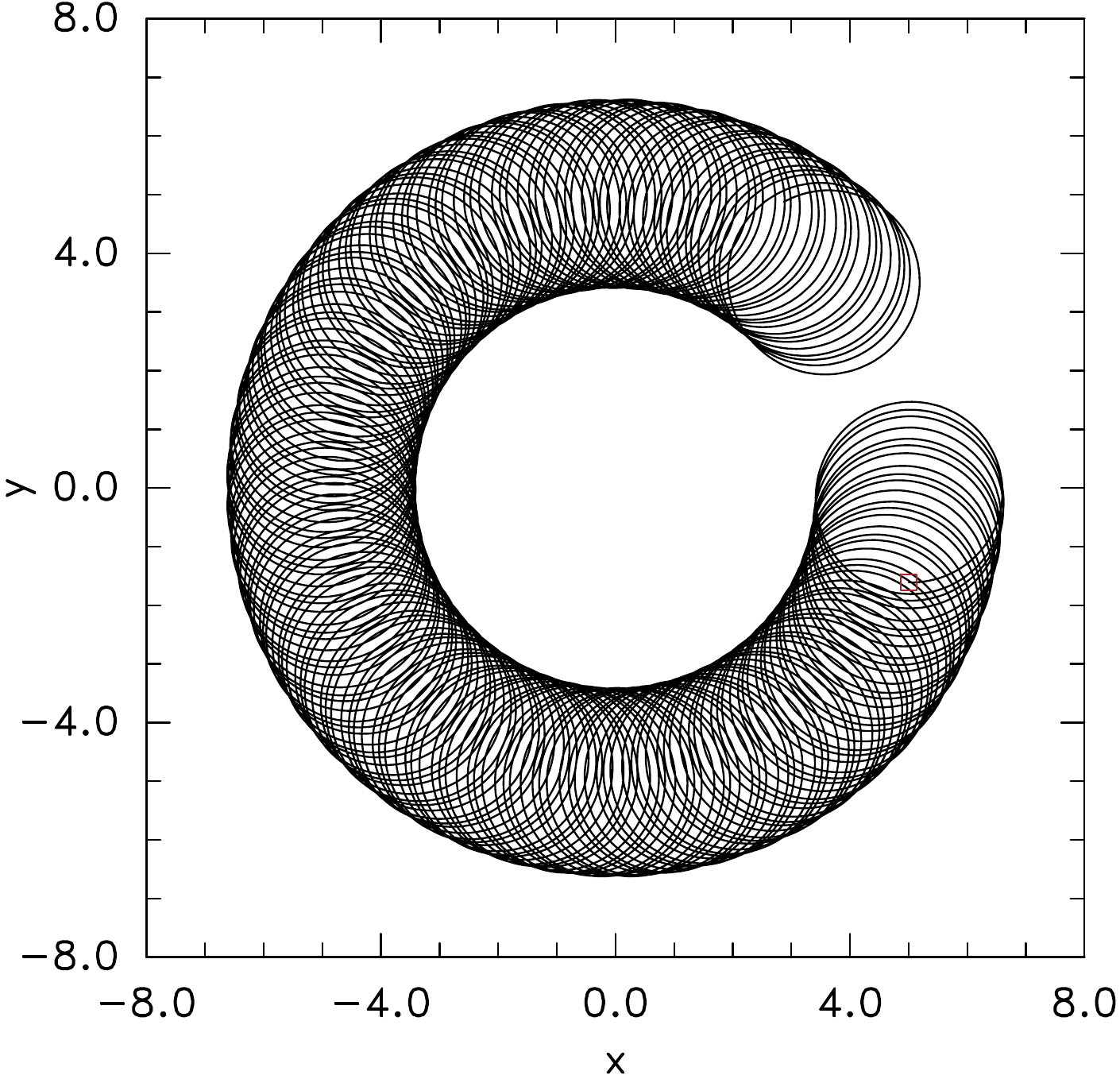}
  \includegraphics[width=0.8\hsize]{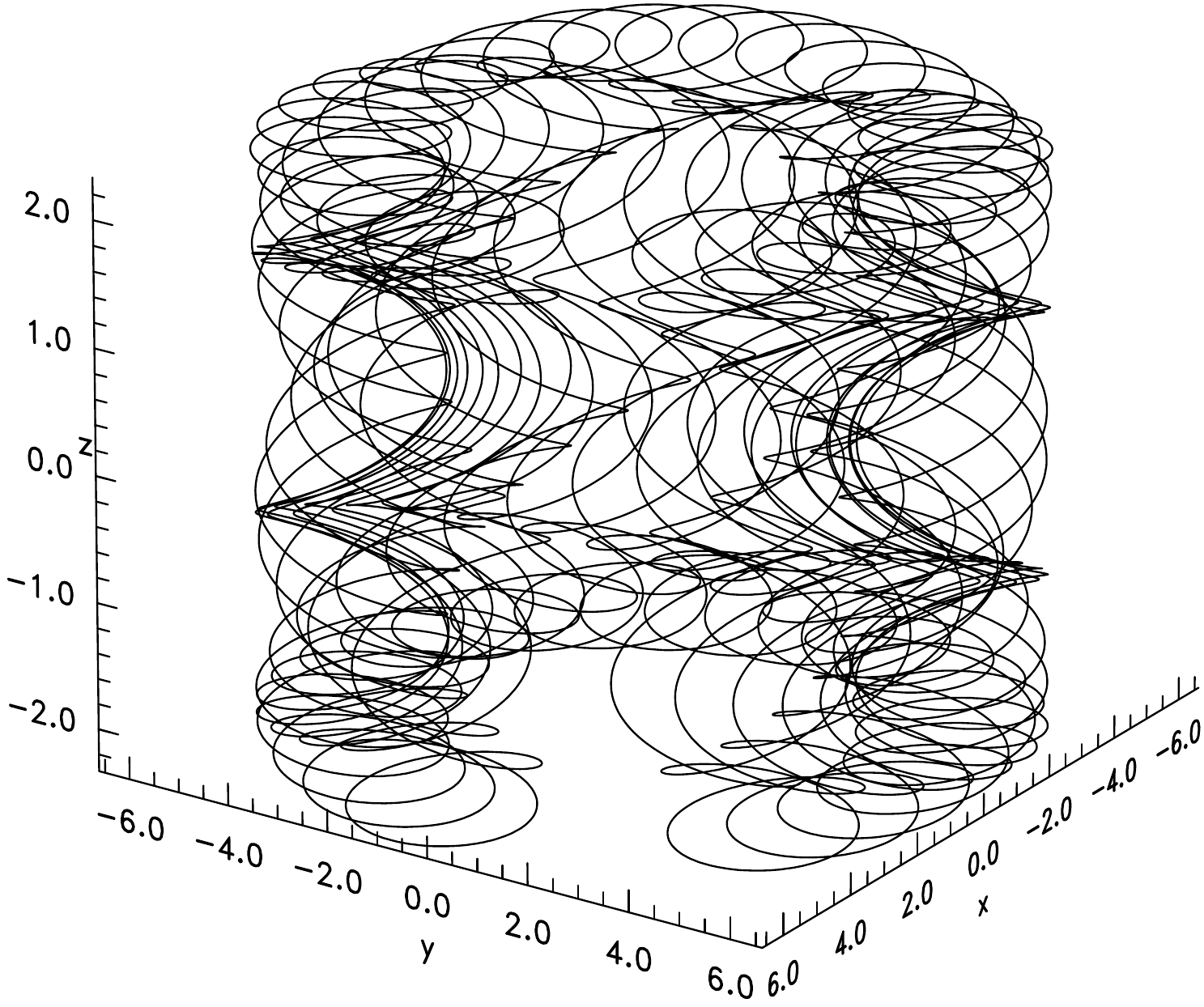}
  \caption{Example of a trapped orbit (a) projected in two dimensions
    showing the axisymmetric direction drift, and (b) in three dimensions
    showing the parallel bounce motion.}
  \label{fig:orbitex1}
\end{figure}
in Fig.\ \ref{fig:orbitex1}. The projection of the orbit on to the
transverse plane shows the drift in the axisymmetric $\theta$
direction of the circular orbit in this situation where the gyro-radius
is a moderate fraction of the guiding-center radius and transverse
potential scale-length. Simultaneously the orbit is bouncing in the
parallel direction, as shown by the three-dimensional perspective.
The shape of the potential in which this particle is moving is given by
the contours of Fig.\ \ref{fig:phiofrz1}.

\begin{figure}
  \centering
  \includegraphics[width=\hsize]{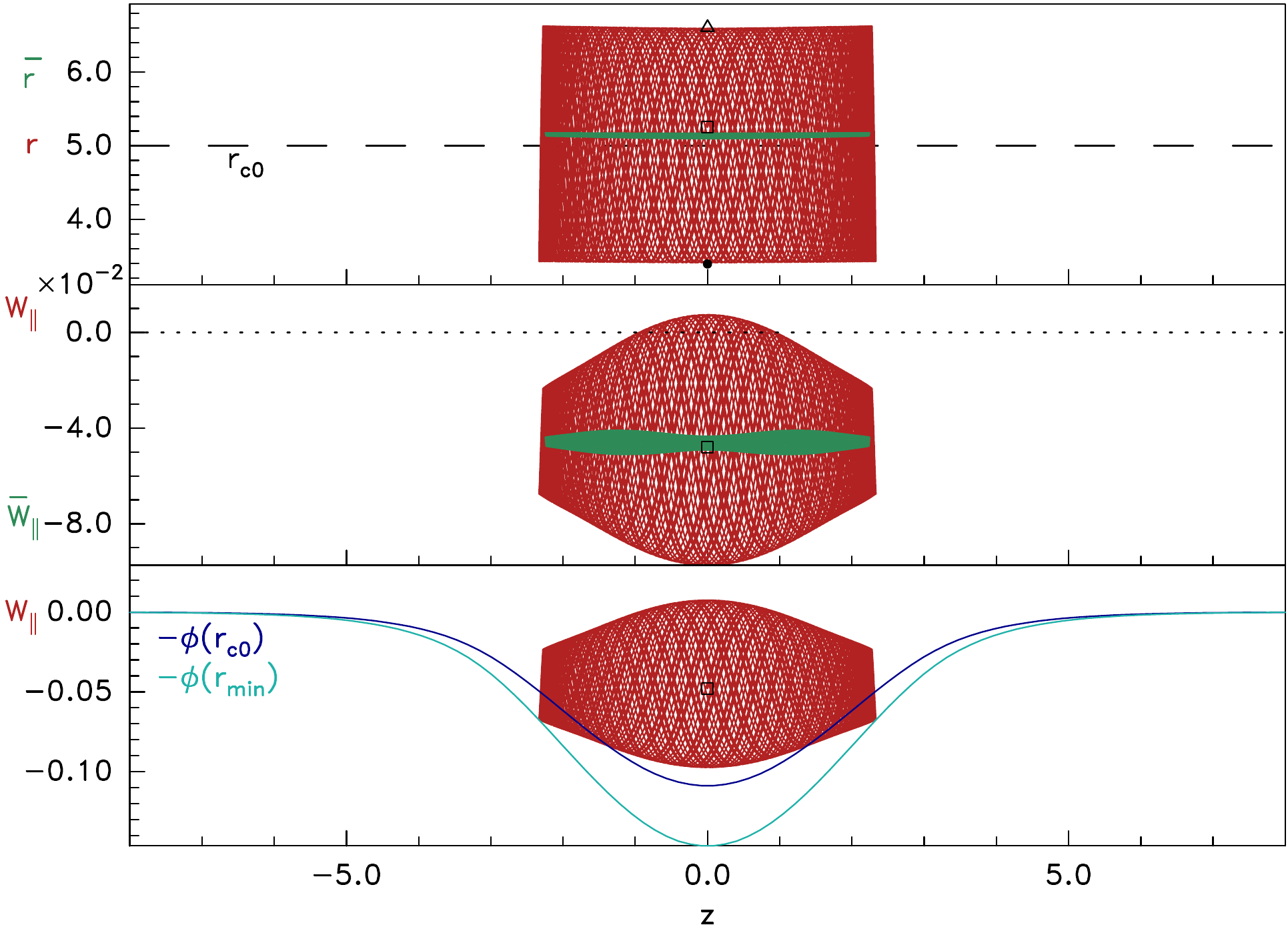}
  \caption{Orbit evolution, showing the particle radius (red) and
    gyro-averaged radius (green) as a function of parallel position
    $z$ (top panel), and the parallel energy $W_\parallel$ (red) and
    its gyro-average $\bar W_\parallel$ (green) together with the
    potential at the guiding-center ($r_{c0}$) and minimum radius
    $r_{min}$ (second and third panels). See text for detailed
    explanation.}
  \label{fig:wpplot1}
\end{figure}

In Fig.\ \ref{fig:wpplot1} is shown the variation of some other
parameters of the orbit.
The top panel shows the instantaneous radial position $r(z)$ (red) and
its single-gyro-period running gyro-average $\bar r$ (green). The
dashed line is the initial radial position of the orbit's
guiding-center (a chosen parameter $r_{c0}$ for this orbit). [Because
there are many bounces of the orbit, individual lines are hard to discern
except at large magnification.] Also shown are the initial particle
position $r_0$ (square) and the positions of the nominal extrema of
the orbit based on the initial gyro-radius, namely $r_{c0}\pm r_{g0}$
(triangle and circle). For this example the gyro-averaged radius
$\bar r$ has very small excursions. On average $\bar r$ slightly
exceeds $r_{c0}$. The initial particle radius exceeds even $\bar r$,
but is controlled by the initial gyro-phase (a chosen parameter),
which in this case is that the initial velocity is directed along the
guiding-center radius from the origin; so $r^2_0=r_{c0}^2+r_{g0}^2$ .

The second panel shows the instantaneous and running gyro-averaged
value of the parallel energy
$W_\parallel=v_\parallel^2/2-\phi(r,z)$. Because of the gyration up
and down the substantial radial potential gradient, the perpendicular
kinetic energy $W_\perp=v_\perp^2/2$ varies considerably with
gyro-angle. Consequently $W_\parallel=W-W_\perp$ varies by an equal
and opposite amount, since $W$ is exactly conserved. (In the code $W$
is monitored and confirms accuracy to a fraction level of
$\sim 10^{-5}$.) Even though $W_\parallel$ can instantaneously exceed
zero, the orbit remains trapped for in excess of 200 parallel bounces.
The gyro-averaged $\bar W_\parallel$ shows some limited variation with
$z$ and gyro-angle, but remains below zero; so the orbit remains
trapped. The lowest panel plots, in addition to $W_\parallel$, the
potential energy at $r_{c0}$ and $r_{min}$. One can observe that the
$z$-extrema of the orbit are approximately where $\bar W_\parallel
=-\phi(r_{c0})$ and $W_{\parallel
  min}=-\phi(r_{min})$ corresponding to zero parallel kinetic energy.

\begin{figure}
  \centering
  \includegraphics[width=\hsize]{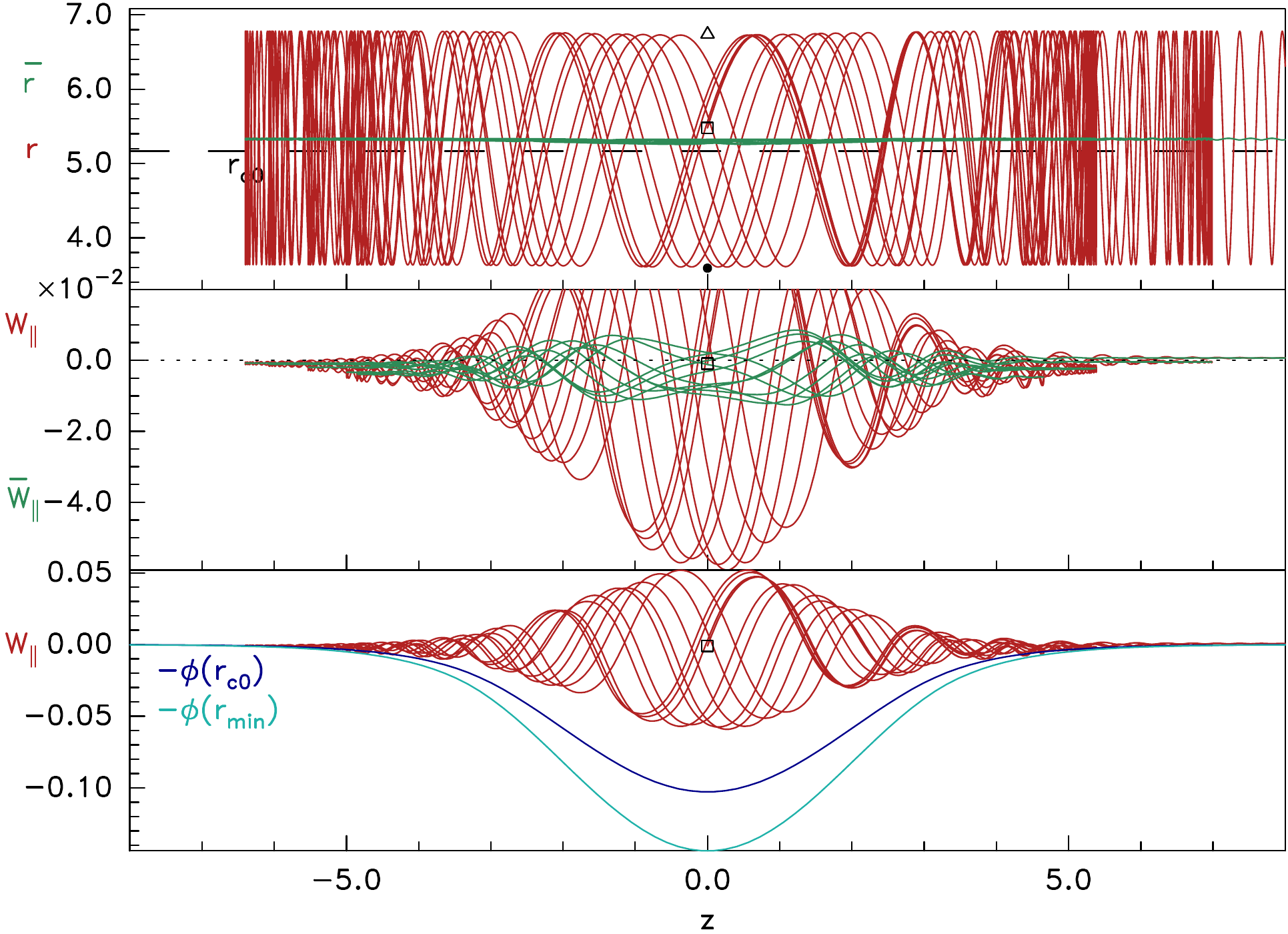}
  \caption{Orbit evolution for a more shallowly trapped orbit than Fig.\
 \ref{fig:wpplot1}, but otherwise the same parameters. }
  \label{fig:wpplot2}
\end{figure}

Fig.\ \ref{fig:wpplot2} shows for comparison an orbit in exactly the
same potential (whose peak is $\psi_0=0.16$) and magnetic field strength
($\Omega=0.9$), and the same total energy ($W=1$), but with a starting
parallel energy that is closer to the trapped-passing boundary
($W_\parallel=-0.13\psi(r_{c0})$ instead of $-0.5\psi(r_{c0})$).
The orbit starts trapped, but only just. Consequently its parallel
excursions extend farther out in $|z|$. Focusing (at high resolution)
on the green line in the second panel $\bar W_\parallel$, one can see
that the first bounce position (at positive $z$) is at $|z|\simeq 4.5$
and $\bar W_\parallel\simeq -0.07\times10^{-2}$. However each
subsequent bounce occurs at an increasing value of $|z|$ because
$\bar W_\parallel$ increases toward zero, until after 10 bounces it
becomes positive and the orbit escapes the hole, moving out to large
positive $z$. In the inner regions of the hole,
$\bar W_\parallel$-variation with gyro-angle is very significant, but
at the ends near the bounce the excursions become small, because the
radial potential gradient is small there, and it is naturally the
value at the ends that determines whether the particle bounces or not.
\begin{figure}
  \centering
  \includegraphics[width=0.9\hsize]{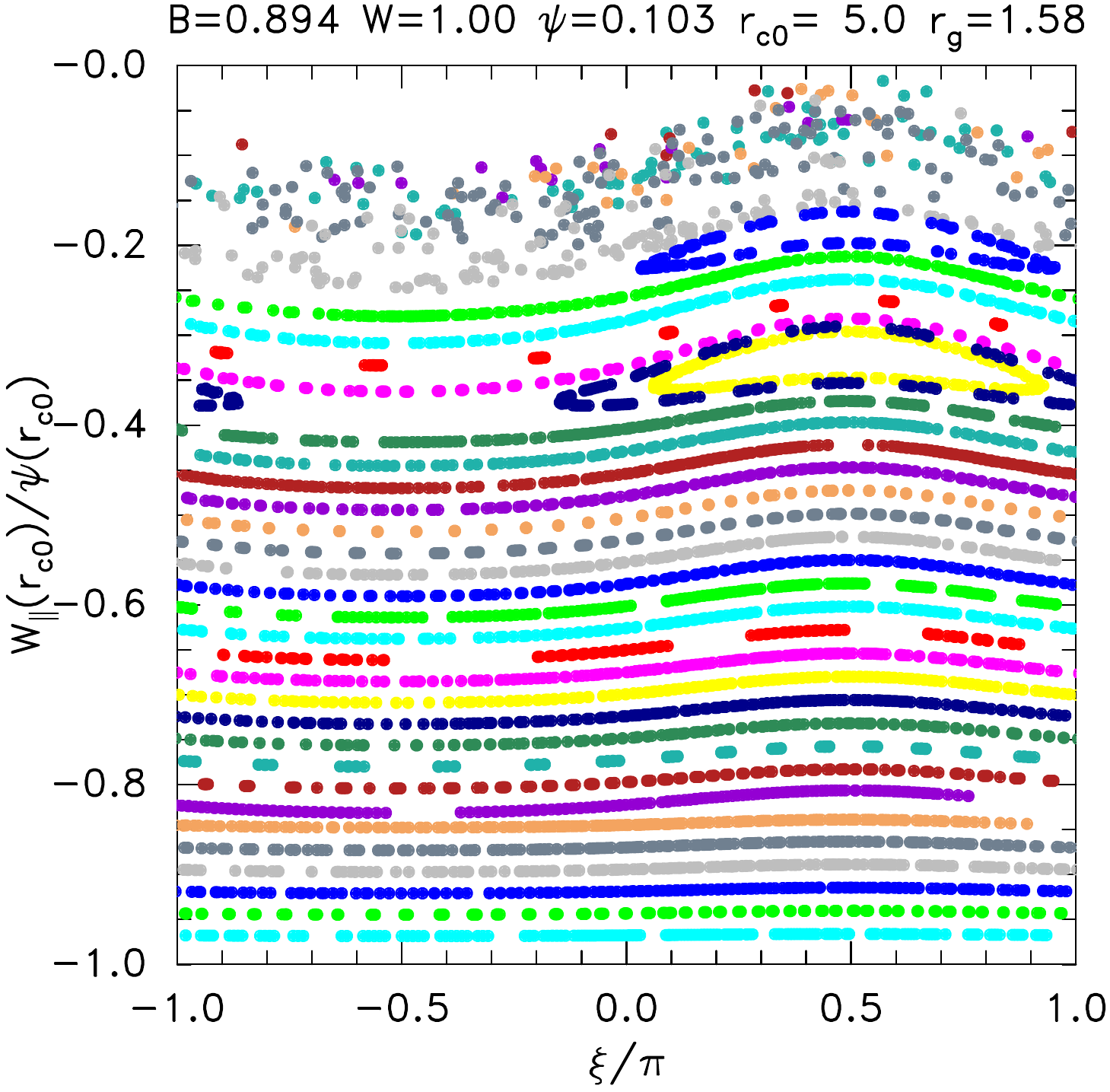}
  \includegraphics[width=0.9\hsize]{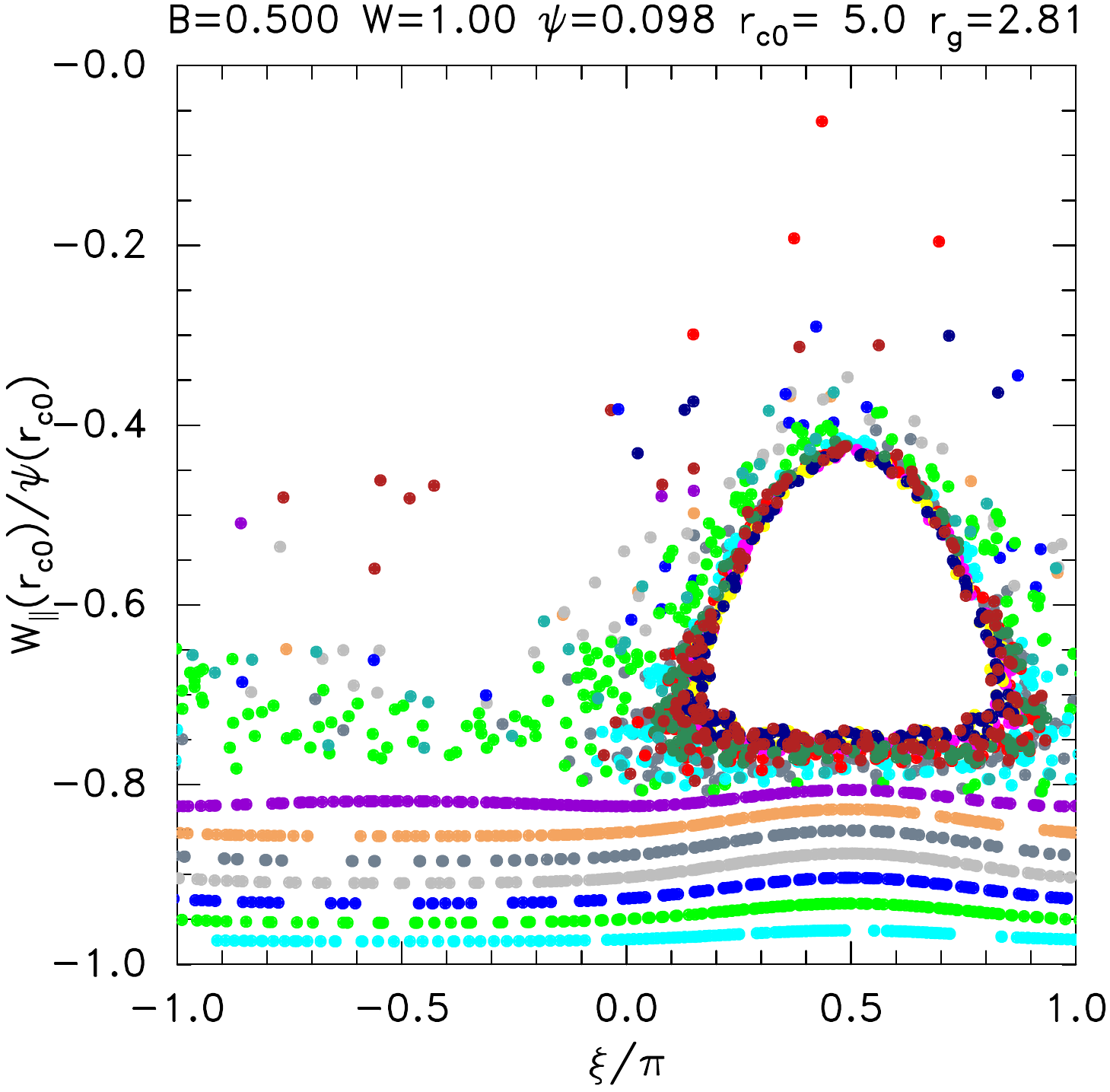}
  \caption{Poincar\'e plot of the parallel energy and gyro-angle at
    passages through $z=0$ for the potential form of
    Fig. \ref{fig:phiofrz1}. (a) for the magnetic field strength
    $\Omega=0.9$ of
    Figs. \ref{fig:orbitex1}, \ref{fig:wpplot1} and \ref{fig:wpplot2},
   (b) for a lower magnetic field, $\Omega=0.5$.}
  \label{fig:poinplot2}
\end{figure}

In Fig.\ \ref{fig:poinplot2}(a) is shown a Poincar\'e plot of the
parallel energy, $W_\parallel(r_{c0})$, evaluated taking the potential
to be that at position $r_{c0}$ (which is a proxy for the
gyro-averaged potential ignoring $\partial^2\phi/\partial r^2$
requiring no averaging), against the gyro-angle for each passage of an
orbit through $z=0$. A sequence of starting $W_\parallel$ orbits is
shown in a corresponding sequence of colors. The orbit of Fig.\
\ref{fig:wpplot1} corresponds to
$W_\parallel(r_{c0})/\psi(r_{c0})\simeq -0.5$. At that energy all the
orbits are permanently trapped and no obvious phase-space islands are
observed. The ratio of the gyro-frequency $\Omega$ to twice the bounce
frequency $2\omega_b$ there is approximately 3.5, so it is not close
to a bounce-cyclotron resonance ($\ell \omega_b = \Omega$ with $\ell$
an even integer). The nearest resonance is $\ell=8$ at
$W_\parallel(r_{c0})/\psi(r_{c0})\simeq -0.35$ where an island is
obvious. By contrast, the orbit of Fig.\ \ref{fig:wpplot2} corresponds
to a starting energy $W_\parallel(r_{c0})/\psi(r_{c0})\simeq -0.13$
which is in the region of apparently stochastic orbits just below the
zero of energy. It corresponds closely to the purple orbit in that
region, which only has a few points because it quickly becomes
detrapped.  Its energy is just above that for the $\ell=12$ and 14
resonances.  Actually the progressive gyro-averaged energy increase
observed in Fig.\ \ref{fig:wpplot2} does not require the orbit
literally to be stochastic in order to become detrapped. But
nevertheless detrapping like this is observed to take place mostly in
proximity to parameters that have overlapped islands and stochastic
orbits. Regions of detrapping (which are also, by time reversal,
regions of trapping) and stochasticity experience strong effective
energy diffusion which will limit any gradients in $f_\parallel$ and
hence suppress $\tilde f$. In this case, a distribution in which
$\tilde f\simeq 0$ down to
$W_\parallel(r_{c0})/\psi(r_{c0})\simeq -0.25$ would be plausible, but
not one that called for substantially non-zero $\tilde f$ closer to
$W_\parallel=0$.

Fig.\ \ref{fig:poinplot2}(b) shows what happens if the magnetic field
strength is lowered but with all else the same. The gyro-resonance
energy is changed so that the $\ell=4$ island centered on
$W_\parallel(r_{c0})/\psi(r_{c0})\simeq -0.6$ comes to dominate the
Poicar\'e plot. In this case making the initial gyro-averaged
potential negative does not represent well the trapping condition,
because almost all energies above the resonance are detrapped.  In
fact all orbits with $W_\parallel(r_{c0})/\psi(r_{c0})> -0.26$ are
lost immediately without even a single bounce, because the
gyro-averaged potential happens to rise above zero in the first
quarter of the bounce period for all such orbits with the standard
starting gyro-phase. At different starting gyro-phase one can find
higher energy orbits with a few (up to 10) bounces prior to loss, but
the overall lost region (whether after zero or a few bounces) of
phase-space is essentially the same. High enough magnetic field avoids
such strong resonances.

\begin{figure}
  \centering
  \includegraphics[width=0.9\hsize]{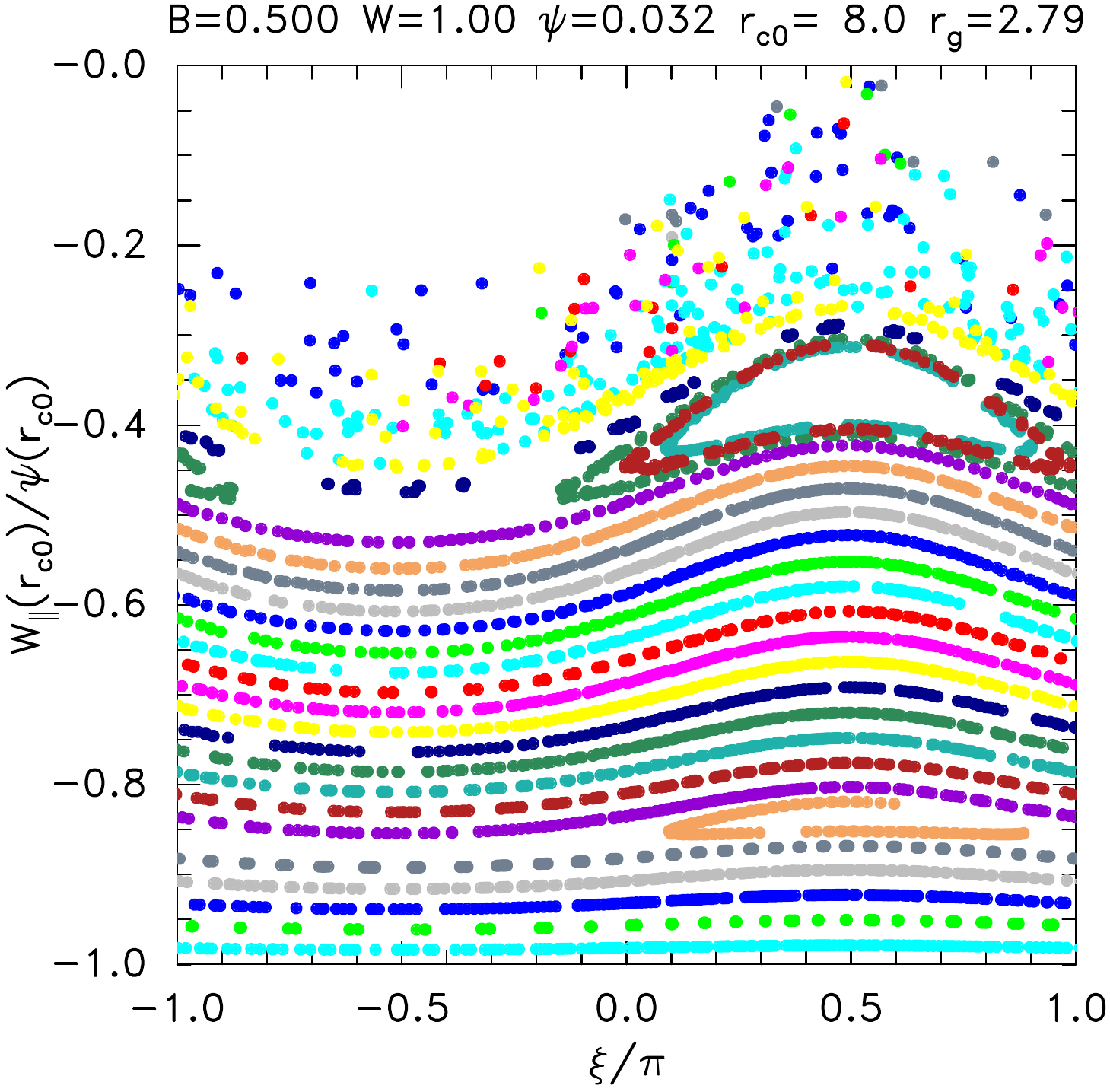}
  \includegraphics[width=0.9\hsize]{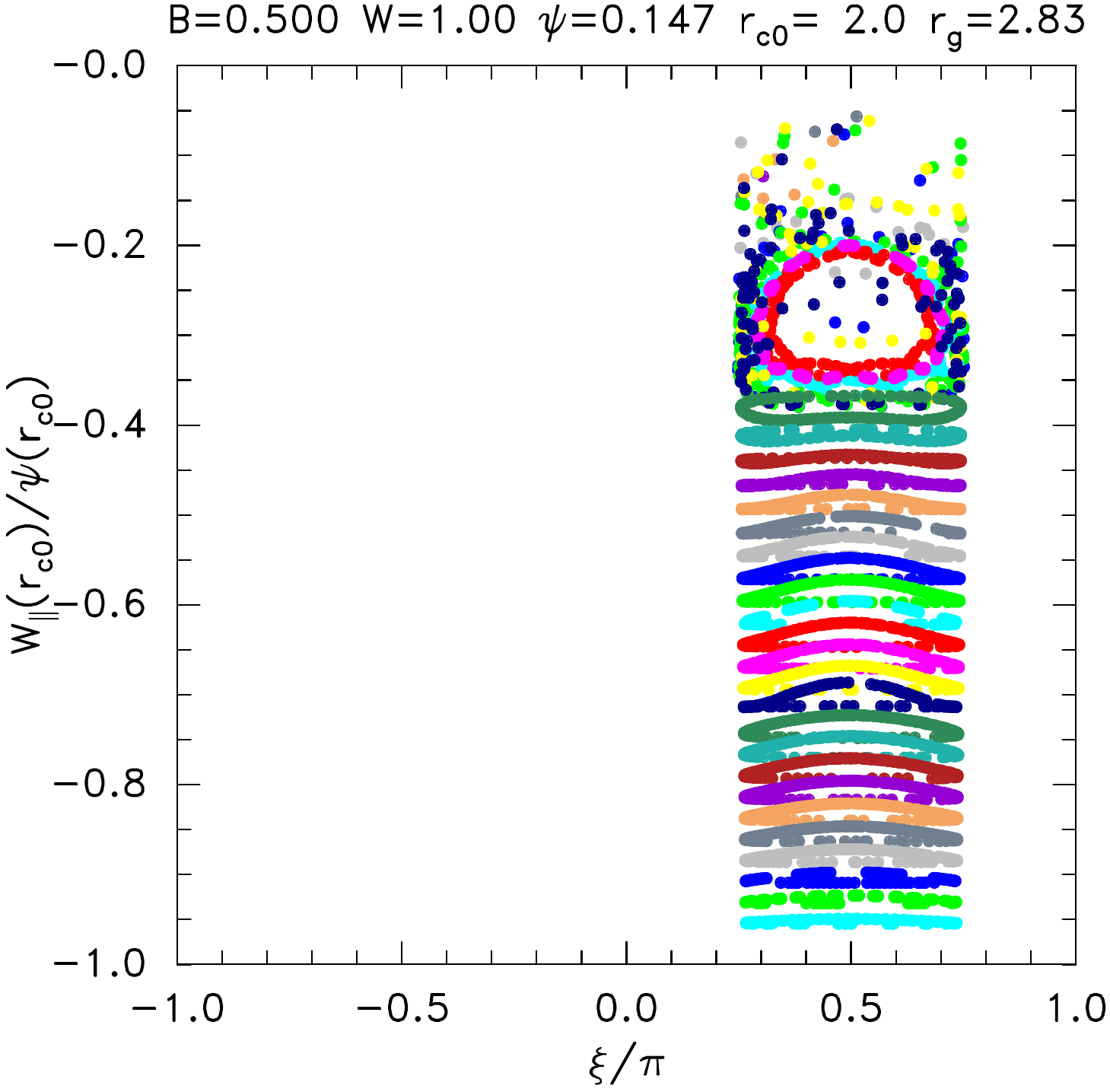}
  \caption{Poincar\'e plots at different guiding-center radii, and the
    lower magnetic field.}
  \label{fig:poinplot4}
\end{figure}

Further trends with guiding-center position are illustrated in Fig.\
\ref{fig:poinplot4}, where different guiding-center position
($r_{c0}$) cases are shown for the same hole parameters as Fig.\
\ref{fig:poinplot2}(b).
Fig.\ \ref{fig:poinplot4}(a) is for a larger guiding-center radial
position $r_{c0}=8$ in the exponentially decaying radial wing of the
hole where the peak potential $\psi$ is substantially smaller. It has
considerably less of its phase-space detrapped than Fig.\
\ref{fig:poinplot2}(b). Fig.\ \ref{fig:poinplot4}(b) shows instead a
case where $r_{c0}<r_g$ so that the gyro-orbit itself encircles the
origin ($r=0$). The resulting excursion of the gyro-phase $\xi$ (which
is defined as the angle between the perpendicular velocity and the
particle radius vector from the origin) is now less than $2\pi$. It
also, though, shows a region of detrapping that is comparable to Fig.\
\ref{fig:poinplot4}(a) and shallower than \ref{fig:poinplot2}(b). Thus
the intermediate radii where $r_{c0}$ lies in the steep radial
gradient of the potential, exemplified by \ref{fig:poinplot2}(b), are
the most susceptible to stochastic detrapping. The extreme limit of
the origin-encircling orbit type is when $r_{c0}=0$, and the orbit
becomes a circle centered on the origin. There are then no resonant
effects and no detrapping because the orbit phase never changes.

What Figs.\ \ref{fig:wpplot2} and \ref{fig:poinplot2} validate is that
combining the gyro-averaged potential energy ($-\bar\phi$) plus the
parallel kinetic energy $v_\parallel^2/2$ gives a quantity
$\bar W_\parallel$ whose sign rather accurately determines whether or
not an orbit bounces; and that, although this quantity has significant
excursions as the orbit moves through the low-$|z|$ strong potential
regions, those do not generally lead to detrapping unless there is
systematic or stochastic enhancement of the $\bar W_\parallel$ at the
bounce positions due to bounce-gyro resonance.

\begin{figure}
  \centering
  \includegraphics[width=.9\hsize]{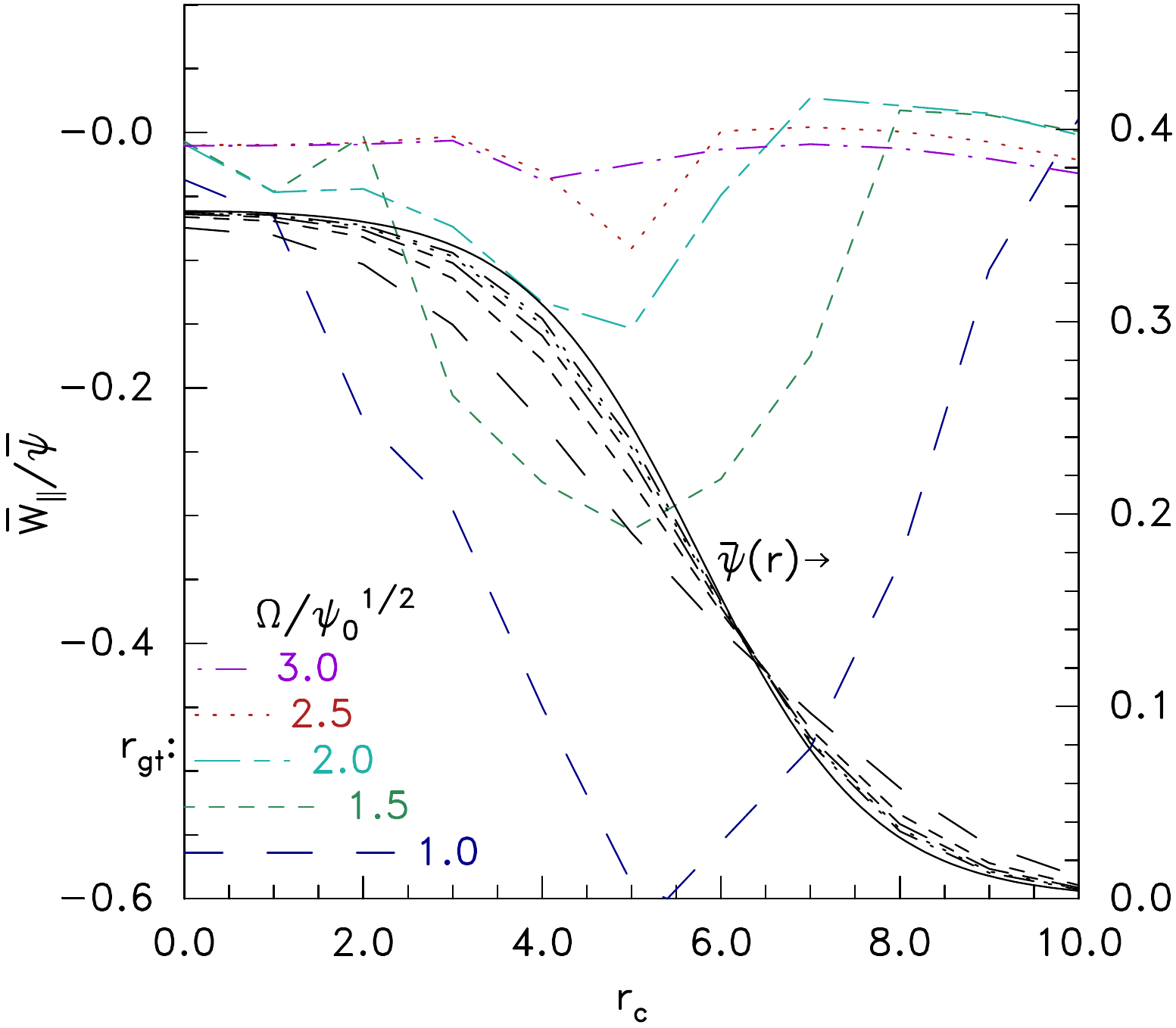}
  \includegraphics[width=.9\hsize]{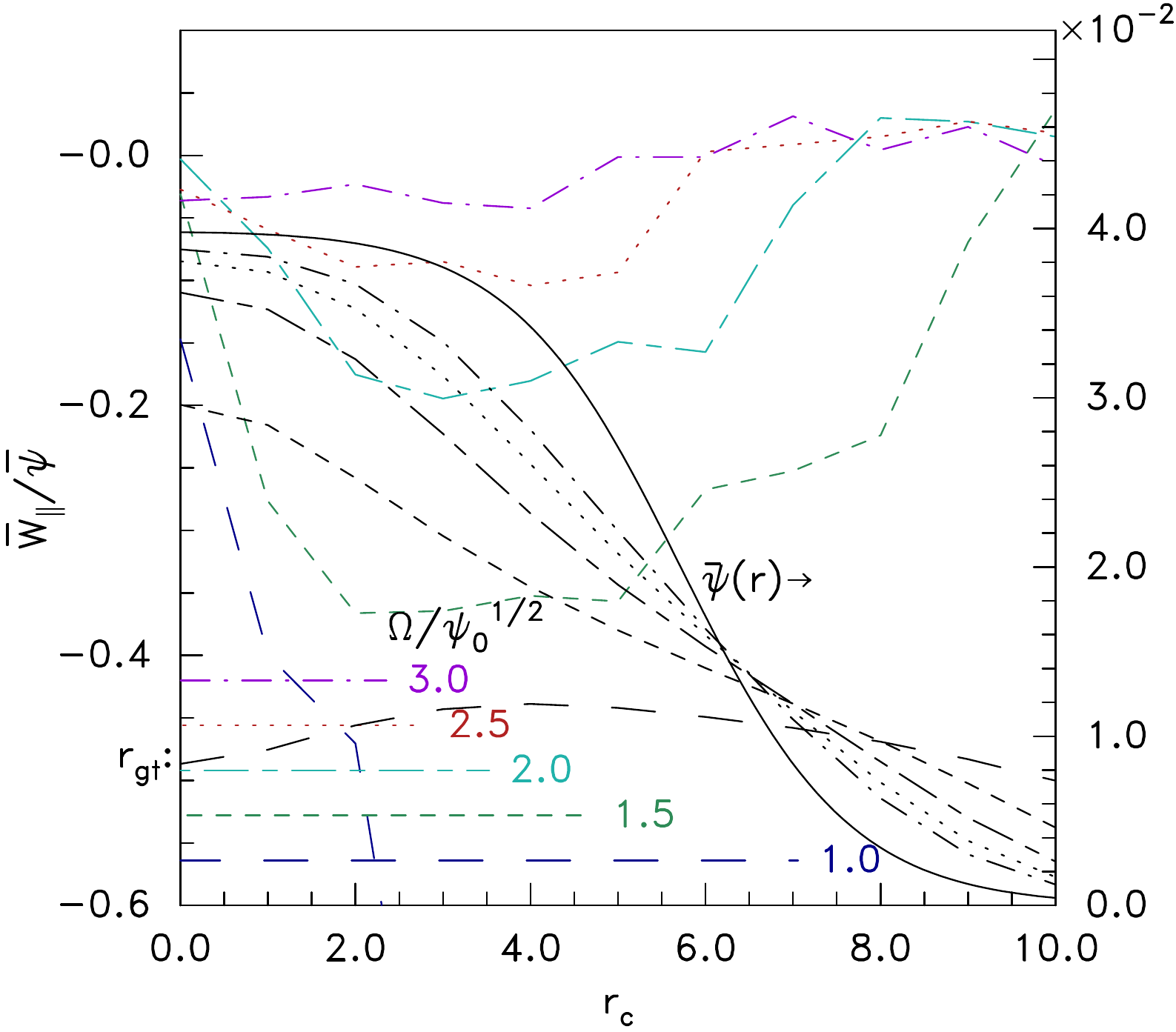}
  \caption{Highest permanently trapped orbit initial parallel energy
    $\bar W_\parallel$ (colors) divided by gyro-averaged potential
    $\bar\psi$, as a function of guiding-center radius for different
    magnetic fields indicated by the length of the thermal gyro-radius
    and the value of $\Omega/\psi^{1/2}$. Also radial profile of
    $\psi$ and $\bar\psi$ (black).  Hole origin potential: (a) $\psi_0=0.36$,
    (b) $\psi_0=0.04$.}
  \label{fig:wpmaxplot}
\end{figure}
This finding is summarized by results for a large number of orbits in
Fig.\ \ref{fig:wpmaxplot}.  It shows as a function of guiding-center
radial position, by colored lines the maximum value of the ratio of
initial parallel energy $\bar W$ to gyro-averaged potential $\bar\psi$
(both at $z=0$) for which the orbit is permanently trapped. Each plot
shows cases for five different magnetic field strengths. That strength
is indicated by the length of the thermal gyro-radius horizontal bars
at the bottom left. It is also given by the labels consisting of the
value of $\Omega/\psi_0^{1/2}$, which is a proxy for $\Omega/\omega_b$
at the hole origin and organizes the results. The radial profile of
the peak potential $\psi(r)$ and the corresponding gyro-averaged
potential $\bar\psi$ (to which $\bar W$ is normalized in
the colored lines) is also plotted in black with the scale on the
right. The different line styles correspond to the different
gyro-radii. Plots have potential at the origin (a) $\psi_0=0.36$ and (b)
$\psi=0.04$, differing by a factor of nine. Yet generally, within the
uncertainty implied by the fluctuations of the curves, they tell the
same story. At high magnetic field, the ratio $\bar W_\parallel/\bar\psi$ is
close to zero, indicating that the gyro-averaged potential
determines parallel trapping over the entire radial profile for all
gyro-radii and $\psi_0$: a key result. This limit is approached when
$\Omega/\psi_0^{1/2}\gtrsim 2$. For lower magnetic fields, detrapping
by gyro-bounce resonance begins at significantly lower
$\bar W_\parallel$, strongest in the central regions of largest
$E_\perp (-d\psi/dr)$, but eventually deep detrapping occurs across most of the
hole radius. Smaller amplitude ($\psi$) holes allow larger gyro-radius
to be accommodated before deep detrapping sets in: Fig.\
\ref{fig:wpmaxplot}(b).

\begin{figure}[htp]
  \centering
  \includegraphics[width=.9\hsize]{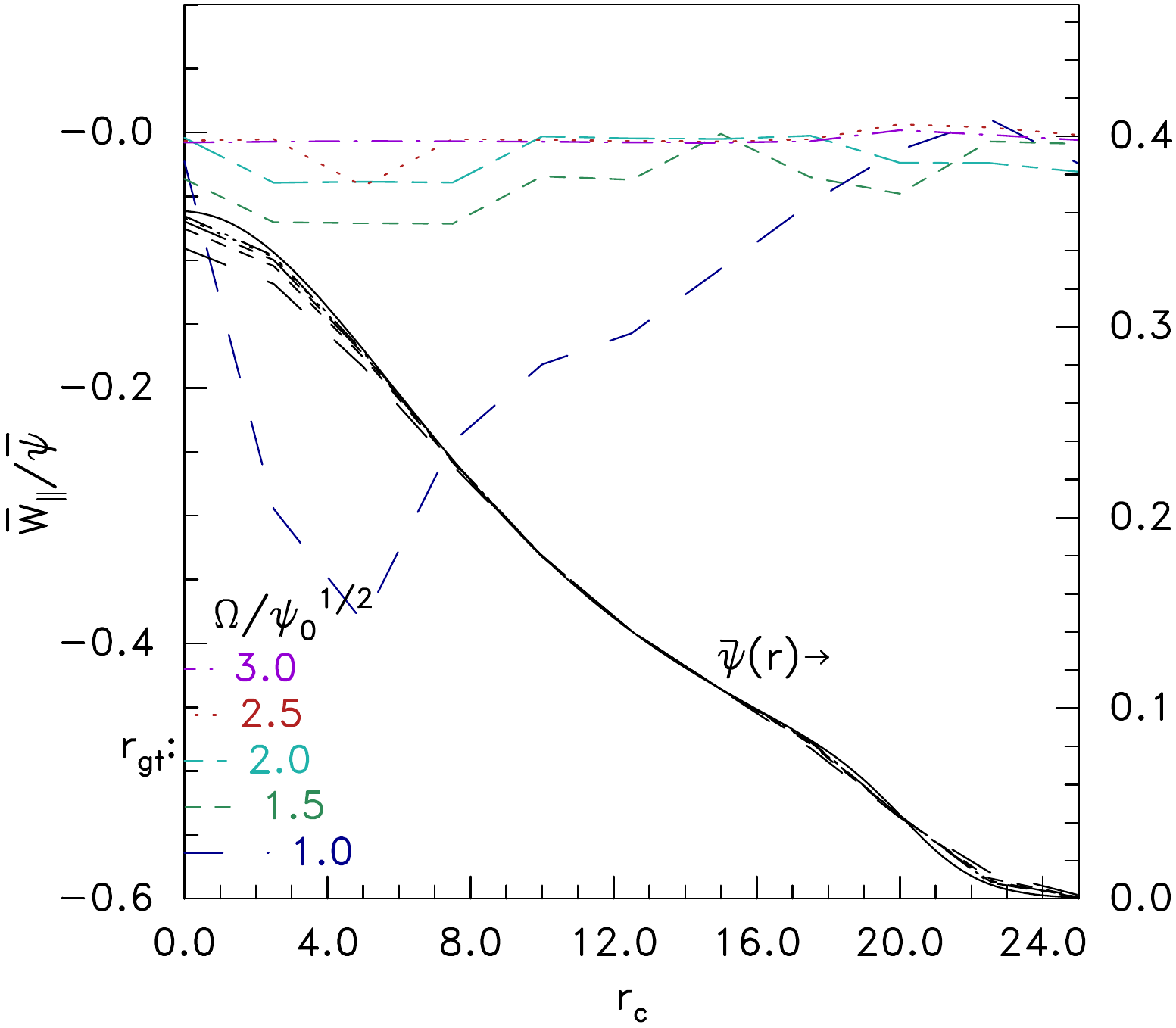}
  \includegraphics[width=.9\hsize]{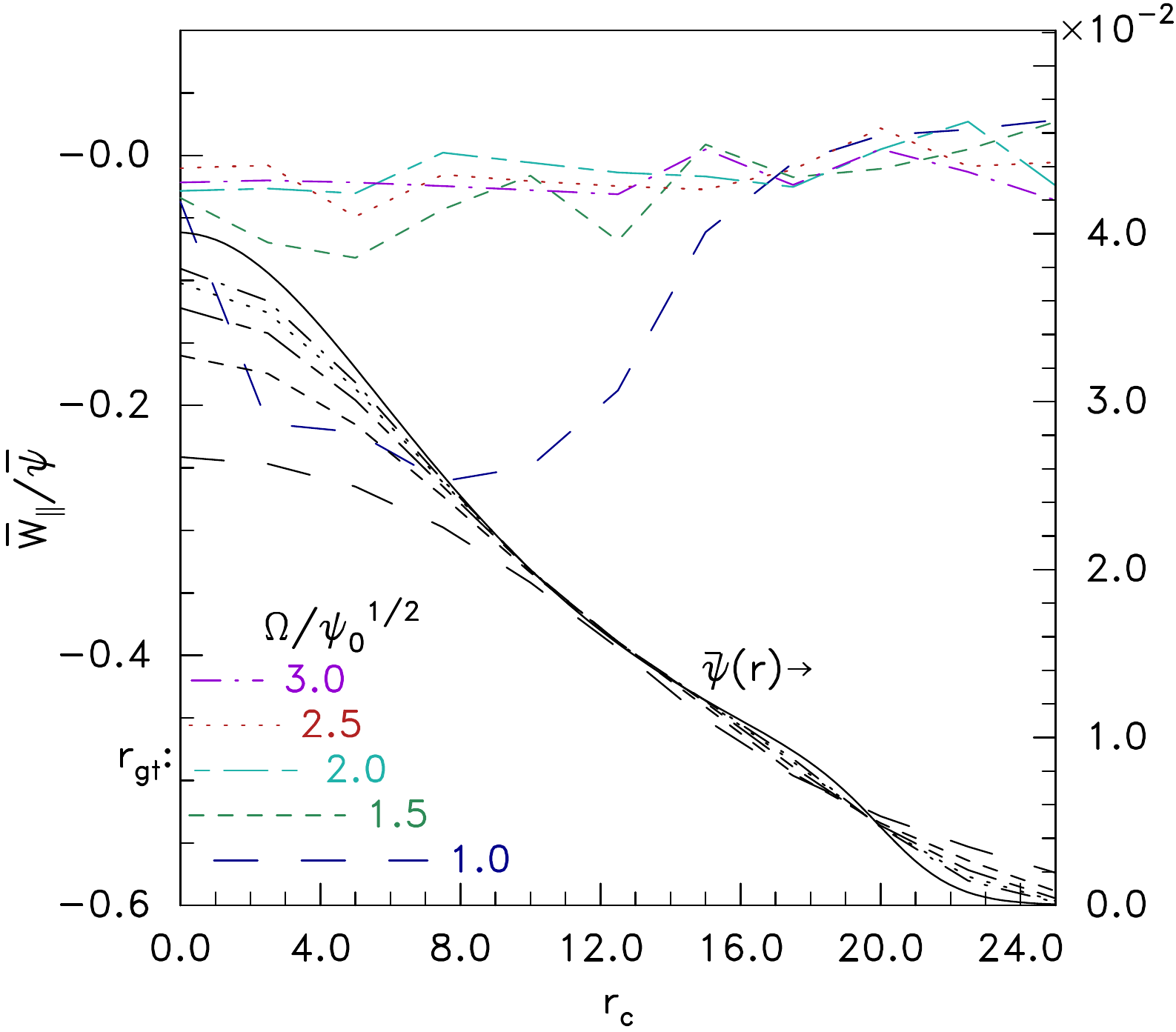}
  \caption{Highest permanently trapped orbit initial parallel energy
    $\bar W_\parallel$ divided by gyro-averaged potential $\bar\psi$,
    as in Fig.\ \protect\ref{fig:wpmaxplot}
    except with a more peaked
    radial profile extending to much larger radius.} 
  \label{fig:wppeakplot}
\end{figure}
Fig.\ \ref{fig:wppeakplot} shows a much wider but more peaked radial
profile case. Detrapping is now strongest relatively close to
the potential peak, which is where the radial field is strongest.  A
threshold below which gyro-bounce resonance causes major detrapping is
still present. It is quantitatively at somewhat lower ratio
$\Omega/\sqrt\psi_0 \sim 1$, in part because of the $E_r$ reduction.

The $\bar\psi$ curves of Figs.\ \ref{fig:wpmaxplot} and \ref{fig:wppeakplot}
illustrate how gyro-averaging is changing the effective potential,
lowering it near $r=0$, and raising it in the positive curvature
wings.

\section{The consequences of gyro-averaging and resonance detrapping}
\label{sec4}
The key consequence of gyro-averaging is that trapped density deficits
become less effective at sustaining the hole potential. That reduced
effectiveness requires deeper deficits in order to sustain the same
potential, leading eventually to a violation of non-negativity of the
distribution function.  It is easiest and most relevant to think of
this in terms of the convolution spreading of the potential profile
given $\tilde n(r)$.

A Gaussian radial potential profile
$\psi(r)=\psi_0{\rm e}^{-(r/a)^2/2}$ provides the simplest
illustration of what gyro-averaging does. When convolved with the
Gaussian of width $\sqrt{2}r_{gt}$ it yields a modified
finite-gyro-radius Gaussian profile $\bar\psi(r)$ of greater width
$\sqrt{a^2+2r_{gt}^2}$.  Of course the total volume
$\int_0^\infty \psi(r)2\pi rdr$ is conserved during convolution,
consequently the height of the smoothed $\bar\psi$ is smaller by the ratio
$\bar\psi_0/\psi_0=(1+2r_{gt}^2/a^2)^{-1}$. This factor gives a strong
reduction of the potential height at the origin once $2r_{gt}$ exceeds
$a$. The same spreading occurs for any shape of profile $\psi(r)$,
when its width is considered to be given by the variance
$\langle r^2\rangle=\int_0^\infty r^2\psi(r)2\pi rdr/\int_0^\infty
\psi(r)2\pi rdr$. For a Gaussian of width $a$,
$\langle r^2\rangle=2a^2$.  For a general profile, when it is
convolved with a Gaussian of width $\sqrt{2}r_{gt}$, its variance
increases as $\langle \bar r^2\rangle=\langle r^2\rangle+4r_{gt}^2$. The
difference is that for a non-Gaussian initial profile its relative
shape changes (toward Gaussian) when convolved, and so the ratio
of central heights $\bar\psi_0/\psi_0$ is not exactly the same, even
though the ratio of the average heights is the same
$(\int \bar\psi 2\pi rdr/\langle \bar r^2\rangle)/(\int \psi 2\pi rdr/\langle
r^2\rangle)=(1+2r_{gt}^2/a^2)^{-1}$. We shall take the average
height suppression to represent the main effect. This amounts to
supposing (correctly) that the inner regions of the hole are the most
important. 

The relative importance of the two main multidimensional electron hole
modifications, gyro-averaging and transverse electric field
divergence, can be estimated by writing the field divergence
$\nabla^2\phi\sim -(a^{-2}+L_\parallel^{-2})\phi$. The potential is
thus suppressed by divergence in a hole of transverse dimension
$\sim a$ relative to a one-dimensional hole with the same charge
density by a factor $F_\perp^{-1}\sim(1+L_\parallel^2/a^2)^{-1}$ where
$L_\parallel$ is a parallel scale length $\sim \lambda_D$ (1 in
normalized units). The
gyro-average potential spreading factor $(1+2r_{gt}^2/a^2)^{-1}$ thus
becomes more important than the effects of transverse electric field
divergence when $r_g\gtrsim\lambda_D$ that is $\omega_p\gtrsim\Omega$ or
$\Omega\lesssim 1$ in normalized units. Since
the threshold of strong depletion by gyro-bounce resonance effects is
$\Omega \sim 2\psi^{1/2}$, a hole depth $\psi$ much smaller than 1
(times $T_e$) will be more strongly suppressed by gyro-averaging than
by transverse divergence, near the gyro-bounce threshold. This point
is made more quantitative in the following.

\subsection{Relationship between electron distribution deficit and potential}

Poisson's equation relates the potential and the electron
distribution (with the ions as an immobile unity background charge density). Integrating it along the
parallel direction requires that
\begin{equation}
  \label{eq:parpois}
  \begin{split}
  0={1\over2}\left.\left(\partial \phi\over \partial z\right)^2\right|_{z=0}&=\int_\infty^0(
  n(\phi)-1- \nabla_\perp^2\phi){\partial\phi\over \partial z} dz\\
  &\approx -\tilde V +(1-\langle\nabla_\perp^2\rangle)\psi^2/2,
\end{split}
\end{equation}
where we have written $-\tilde V\equiv\int\tilde nd\phi$,
$\langle\nabla_\perp^2\rangle\equiv2\int\nabla_\perp^2\phi
d\phi/\psi^2$, and $n(\phi)=n_f+\tilde n $ is the sum of a reference
distribution that is independent of parallel velocity in the trapped
region and $\tilde n$ is the deficit of the trapped electron density
with respect to it. The final equality of eq.\ (\ref{eq:parpois})
approximates $n_f\simeq 1+\phi$ and approximates the potential form as
separable.  $\nabla_\perp^2\phi/\phi$ is then a function only of $r$
and equal to $\langle\nabla_\perp^2\rangle$ which can be taken out of
the $z$-integral; for non-separable potential it must remain in
integral form. Since $\tilde n =\int_0^{\sqrt{2\phi}} \tilde f dv$,
the classical potential $\tilde V$ scales like
$\sim -\tilde f \psi^{3/2}$, and eq.\ (\ref{eq:parpois}) means
$-\tilde f\sim (1-\langle\nabla_\perp^2\rangle ) \psi^{1/2}$. This
scaling is universal.

In the ``power deficit model'', worked out in detail
elsewhere\cite{Hutchinson2021a}, the trapped parallel distribution
deficit $\tilde f$ is presumed to be of the specific form
\begin{equation}
  \label{powerdef}
  \tilde f(W_\parallel) = \tilde f_\psi\left(W_j-W_\parallel\over W_j+\phi\right)^\alpha
\end{equation}
varying between zero at parallel energy
$W_\parallel=v_\parallel^2/2-\phi = W_j$ and $\tilde f_\psi$ at
parallel energy $W_\parallel=-\psi$ (the bottom of the potential well). Fractional
powers of negative quantities are by convention taken to be zero, so
there is zero deficit for energies $W_\parallel >W_j$, i.e. near the
trapped-passing boundary $W_\parallel =0$. In the absence of
gyro-averaging, the condition (\ref{eq:parpois}) on the classical
potential $V=\int \rho {d\phi\over dz}dz$ for the power deficit model
has analytic coefficients and becomes:
\begin{equation}
  \label{fpbc}
  -\tilde f_\psi\psi^{3/2} (1+W_j/\psi)^{3/2} {2 G\over \alpha+3/2} = -F_\perp
  V_f(\psi)\simeq F_\perp \psi^2/2.
\end{equation}
Here $G=\sqrt{\pi/2}\Gamma(\alpha+1)/\Gamma(\alpha+3/2)$ is a constant
depending only on $\alpha$, expressible in terms of the Gamma
function, $V_f(\phi)$ is the classical potential of the flat-trapped
screening density $n_f\simeq 1+\phi$ minus the background uniform ion
density, and $F_\perp$ is the local correction factor
$F_\perp(r)=(1-\langle\nabla_\perp^2\rangle)$ that accounts for
perpendicular electric field divergence in Poisson's equation. Now
$F_\perp$ and $V_f(\phi)\simeq-\phi^2/2$ are unaltered by
gyro-averaging because they depend on the local, not gyro-averaged,
potential\cite{Hutchinson2021}. But the left hand side term
$\tilde f_\psi(\psi+W_j)^{3/2}$ represents the trapped electron
guiding-center distribution deficit integrated with respect to
$\bar\phi$, and must account for gyro-averaging. So in the presence of
finite gyro radius it must be taken as
$\tilde f_{c\bar\psi}(\bar\psi+\bar W_j)^{3/2}$, and on average
$\bar\psi/\psi \simeq (1+4r_{gt}^2/\langle r^2\rangle_0)^{-1}$. We
will suppose that $\bar W_j/\bar\psi=W_j/\psi$ is independent of
gyro-averaging. If instead $\bar W_j=W_j$ were kept constant as
$\psi$ was averaged, the constraints about to be discussed would be
made more severe.

To compensate for the reduction in trapped energy range caused by
$\bar\phi$ averaging reduction, if we wished to preserve $\psi(r)$
while increasing $r_{gt}$, the condition (\ref{fpbc}) then
requires a greater trapped guiding-center deficit
\begin{equation}
  \label{fpsibar}
  \tilde f_{c\bar\psi} = \tilde f_\psi (1+4r_{gt}^2/\langle r^2\rangle)^{3/2}.
\end{equation}
The resulting cubic dependence of $|\tilde f_{c\bar\psi}|$ on $r_{gt}$,
if $r_{gt}^2/\langle r^2\rangle$ is allowed to increase significantly
beyond unity by lowering the magnetic field strength, will soon cause
the trapped electron deficit to exceed the allowed maximum value
(0.399) permitted by non-negativity of the total guiding-center distribution
function. Non-negativity therefore constrains ${\langle r^2\rangle}$
not to be very much smaller than $r_{gt}^2$. An alternative way to
accommodate gyro-averaging increase might be to decrease $\psi$,
thereby reducing the required $|\tilde f_{c\bar\psi}|$. However, if the
magnitude of $|\tilde f_{c\bar\psi}|$ is fixed by non-negativity, then
eq.\ (\ref{fpsibar}) requires
$\psi\sim (1+4r_{gt}^2/\langle r^2\rangle)^{-3}$, an even stronger
(sixth power) dependence on gyro-radius, which would rapidly force the
hole potential to become negligible.  We can conclude that electron
hole equilibria exist in the presence of gyro-averaging only if
\emph{increase of gyro-radius is accompanied by increase of transverse
  potential extent approximately keeping pace with gyro-radius}. Or
expressed more briefly: $4r_{gt}^2\lesssim \langle r^2\rangle$.
Although we have demonstrated this effect for a particular model of
the parallel distribution, the scalings, if not the precise
coefficients, of all the phenomena are virtually independent of the
model or the transverse profiles.

It should be remarked that the empirical scaling of Franz et al,
$L_\perp/L_\parallel\sim \sqrt{1+(\omega_p/\Omega)^2}$, is in the
present terminology $\langle r^2\rangle=L_\parallel^2
(1+1/\Omega^2)=L_\parallel^2(1+r_{gt}^2)$, and taking $L_\parallel=2$
is indistinguishable (within their significant observational
uncertainty) from $ \langle r^2\rangle= 4r_{gt}^2$.

\subsection{Allowable electron hole parameter space}
\begin{figure}
  \centering
  \includegraphics[width=0.9\hsize]{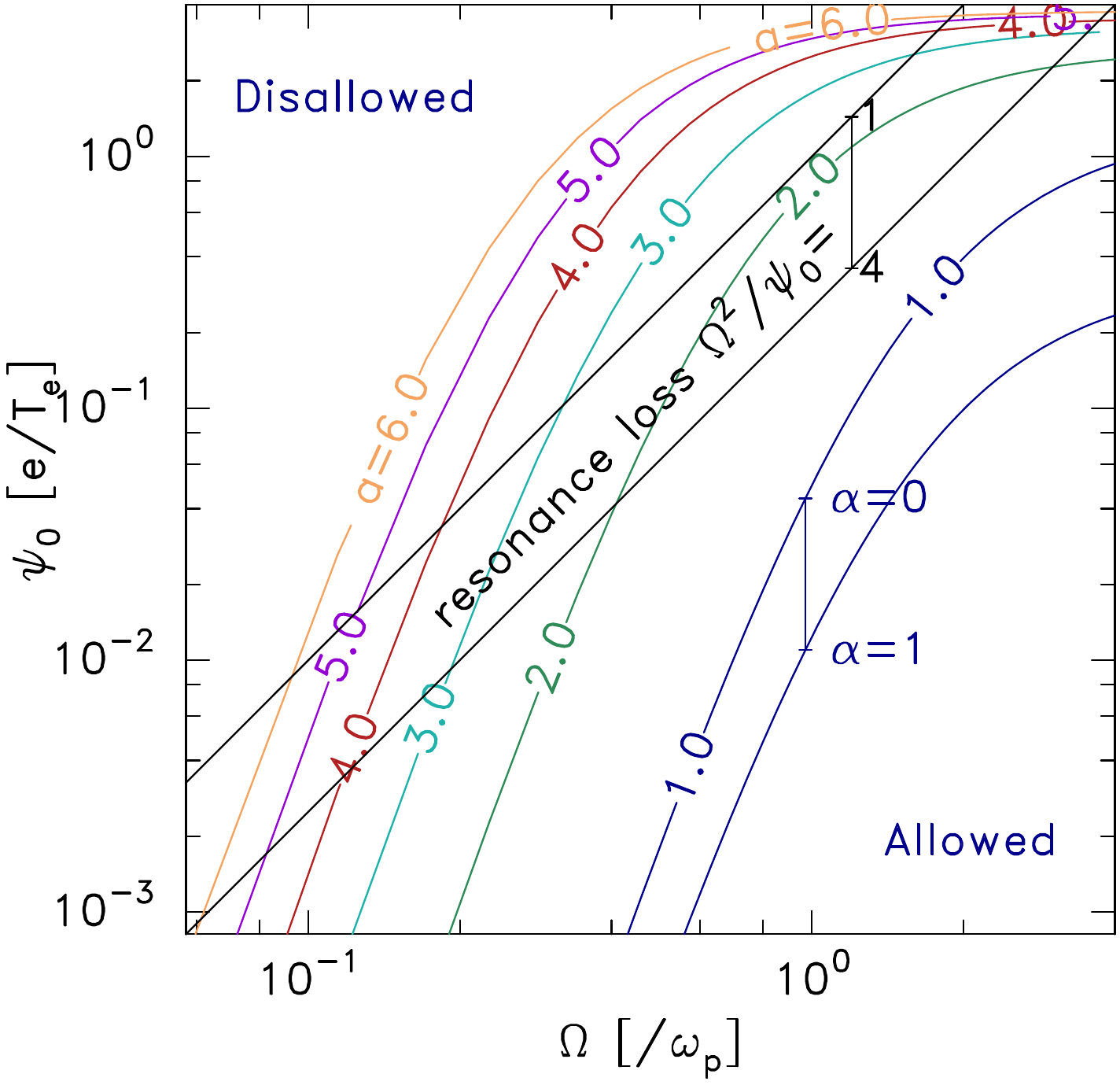}
  \caption{The limits of allowable peak potential permitted by
    non-negativity of $f$ (colors) and avoidance of gyro-bounce
    resonant detrapping (black). Non-negativity depends on the radial
    width $a=(-\langle\nabla_\perp^2\rangle_0/2)^{-1/2}$. It also
    varies depending on the parallel-peaking coefficient $\alpha$;
    curves are highest for a waterbag deficit $\alpha=0$. The range is
    indicated by the vertical bar. The approximate uncertainty range
    of the resonance loss is indicated similarly. The parameters must
    lie below both relevant lines.}
  \label{fig:psiomegalim}
\end{figure}

Let us write equation (\ref{fpbc}) at the origin (the most demanding place)
with the maximum allowable $-\tilde f_{c\bar\psi_0}=1/\sqrt{2\pi}$ to
avoid a negative distribution function. Since we are focussing on $r=0$,
take the transverse potential shape there to be approximately Gaussian
$\propto{\rm e}^{-(r/a)^2/2}\simeq 1-(r/a)^2/2$ with width
$a=\langle r^2\rangle^{1/2}_0/\sqrt{2}$ and
$ \langle\nabla_\perp^2\rangle_0=-2/a^2$, so the divergence and
gyro-averaging factors are $(1+2/a^2) [1+2/(\Omega^2a^2)]^{3/2}$. Then
\begin{equation}
  \label{fmax}
  \psi_0^{1/2}={(1+W_j/\psi)^{3/2}
    \over (1+2/a^2) [1+2/(\Omega^2a^2)]^{3/2}}
  {2 G\over\sqrt{\pi/2}( \alpha+3/2)}.
\end{equation}
Now at the maximum permissible value of $(1+W_j/\psi)$, the
coefficient $(1+W_j/\psi)^{3/2}2G/(a+3/2)$ is approximately 2.4 at
$\alpha=0$ and falls to about 1.2 at $\alpha=1$ because of the
$z$-peaking of the profile. If we adopt the $\alpha=0$ value (the most
forgiving) and write
$2.4/\sqrt{\pi/2}\simeq 2$ for the total coefficient we obtain for the
maximum allowable potential
\begin{equation}
  \label{eq:psiofomega}
  \psi_0\simeq 4(1+2/a^2)^{-1} [1+2/(\Omega a)^2]^{-3}.
\end{equation}
This gives the maximum of $\psi_0$ versus $\Omega$
for some chosen $a$. Lines below which $\psi_0$ must lie are shown for a
range of widths $a$ in Fig.\ \ref{fig:psiomegalim}
For $\alpha=1$ the lines shift downward by a factor $\sim4$ as shown
by the bar and associated line for $a=1$. Permissible equilibria lie
below the lines.  Obviously points for which $-\tilde f_\psi$ is
actually less than $1/\sqrt{2\pi}$, so the total trapped distribution
is greater than zero, lie below the line by an additional factor, which is
$-\tilde f_\psi^22\pi$.

The other major constraint on equilibrium is the detrapping of
electrons by gyro-bounce resonance. Based on the prior studies of
exponential potential gradients\cite{Hutchinson2020},
$\psi_0$ must lie below $\psi_0\sim \Omega^2/4$ to avoid strong
detrapping. The explorations of section \ref{orbitint} (Fig.\
\ref{fig:wppeakplot}) show that for wider holes with lower peak
gradients the limit is closer to
$\psi_0\sim \Omega^2$. These limit lines give the typical range of resonance
loss thresholds, shown by two black straight lines in Fig.\
\ref{fig:psiomegalim}. The strong detrapping effects take
place not at $r=0$ but in the steep region of the potential profile;
even so, they prevent a steady equilibrium shape having $\psi_0$ above
the threshole, and actually tend to shrink the transverse width, making the
losses worse and tending to collapse the hole entirely.

\section{Discussion}

The theory described here has addressed the three main effects of
finite transverse extent on electron hole equilibria (1) transverse
electric field divergence, (3) orbit detrapping by gyro-bounce
resonance, and (4) gyro-averaging effects on the potential and density
deficit. [Supposed modification of the shielding, enumerated (2) in
the introduction, does not occur.] It has explored a sampling of
different self-consistent transverse shapes.

In addition to the plausibility constraint that the distribution
deficit $\tilde f$ should be zero at and immediately below the
trapped-passing boundary, the two critical physics constraints of
non-negativity and avoidance of resonant detrapping have been applied
to obtain bounds relating the peak potential $\psi_0$, magnetic field
strength $\Omega$ (equivalent to the inverse of the thermal
gyro-radius $r_{gt}$), and the transverse extent
$a\ (= \sqrt{\langle r^2\rangle/{2}})$. Fig.\ \ref{fig:psiomegalim} shows
these limits graphically.

The convenient specialization of the current work to axisymmetric
geometry has avoided complications associated with fully three
dimensional structures. The principles of gyro-averaging of potential
and of particle deficit, and the principles of gyro-bounce orbit
detrapping, will apply fully regardless of non-axisymmetry. Since
gyro-orbits remain circular in the drift frame, their averaging
remains isotropic in the transverse plane, and the effective
adjustment depends to lowest order on the transverse Laplacian of the
unaveraged potential. The gyro-bounce detrapping perturbation depends
on the potential through the \emph{magnitude}, not direction, of its
gradient. Since in the wings Debye shielding gives rise to
$|\nabla_\perp\phi/\phi|\simeq 1$ it is expected to have similar
behavior there regardless of non-axisymmetry. The central regions
expected to be most susceptible to resonant detrapping would be those
where the transverse gradient is largest. However, the dependence of
detrapping on $E_\perp$ is weaker than its dependence on
$\Omega/\sqrt\psi$ until $E_\perp$ becomes very small. Overall, since
Fig.\ \ref{fig:psiomegalim} is independent of hole geometry apart from
the parameter $a$, whose inverse equals the square root of the
normalized Laplacian ($\sqrt{\nabla_\perp^2\phi/\phi}$) at the origin, it may be expected to apply to
non-axisymmetric electron holes to approximately the same degree that
it applies to axisymmetric ones.

It should be recognized that while non-negativity and Poisson's
equation are instantaneous requirements, resonant detrapping takes a
significant time to deplete the trapped deficit: typically some
moderate number of bounce-times, and certainly at least $\sim$ the
gyro-period.  Therefore, if a multidimensional electron hole is formed
approximately in a bounce-time or faster, for example by a highly
nonlinear bump-on tail instability or some subsequent process such as
transverse break-up of an initially one-dimensional hole, then it
takes a longer time for the resonant detrapping to become
important. It might then be possible to observe multidimensional
electron holes shortly after their formation that violate the resonant
detrapping criterion.  Indeed, avoiding all resonance loss is such a
severe requirement that it prevents the long term sustainment of
multidimensional electron holes of any significant potential amplitude
(e.g. $\psi_0>10^{-2}T_e/e$) below magnetic field strength
corresponding to $\Omega/\omega_p\sim 0.1$. The present results thus
suggest that any solitary potential structures observed in space
plasmas where $\Omega/\omega_p\lesssim 0.1$ are either some phenomenon
different from electron holes, or they are just-formed electron holes
that will quite soon disappear by resonant detrapping.

The gyro-averaging effects combined with non-negativity act in a
shorter time. Above the green curves of Fig.\ \ref{fig:psiomegalim},
orbits escape in approximately a quarter of a bounce period. This
duration is so short that such a situation, violating the constraint,
cannot be considered a Vlasov equilibrium, because $f$ is not a
function of energy. It is also much less likely that such an object
would be observed, because of its short life. If then the resonance
loss constraint (black curve) was violated for some moderate time
duration, but the non-negativity constraint was not, and became the
determining factor, then the simplified theoretical scaling
$ \langle r^2\rangle\gtrsim 4r_{gt}^2$ would be expected. Equality in
this equation is indistinguishable from the empirical scaling of Franz
et al\cite{Franz2000}. Nevertheless the experimental proxy they used
for aspect ratio ($E_\parallel/E_\perp$) has been
shown\cite{Hutchinson2021a} to be an extremely uncertain measure of
$L_\perp/L_\parallel$; so one should not make too much of this
agreement with observations. A great deal remains to be done to
establish observationally what the structure of multidimensional holes
in space actually is. One can look forward to a critical comparison of
the theory presented here with future observations.

\subsection*{Acknowledgements and Supporting Material}
I am grateful to Greg Hammett for helpful discussions of the
foundations of gyrokinetics, to David Malaspina for insights into the
instrumentation and data of satellite measurements, and to Ivan Vasko
for discussions of electron hole interpretation of space
observations. The code used to generate the equilibria is available at
\href{https://github.com/ihutch/helmhole}{https://github.com/ihutch/helmhole}, and the orbit-following code at \href{https://github.com/ihutch/AxisymOrbits}{https://github.com/ihutch/AxisymOrbits}. Figures were generated by
these codes or the explicit equations in the article, not from separate data.
No public external funding for this research was received.

\bibliography{JabRef}

\end{document}